\documentstyle[amsmath,amssymb,aps,epsf,eqsecnum,graphicx]{revtex}

\begin{document}

\title{Monge Metric on the Sphere \\
and Geometry of Quantum States}
\author{Karol {\.Z}yczkowski$^{1,2}$ and Wojciech S{\l}omczy{\'n}ski$^3$}
\address{$^1$Centrum Fizyki Teoretycznej, Polska Akademia Nauk, \\
Al. Lotnik{\'o}w 32, 02-668 Warszawa, Poland}
\address{$^2$Instytut Fizyki im. Mariana Smoluchowskiego, \\
Uniwersytet Jagiello{\'n}ski, ul. Reymonta 4, 30-059 Krak{\'o}w, Poland}
\address{$^3$Instytut Matematyki, Uniwersytet Jagiello\'nski, \\
ul. Reymonta 4, 30-059 Krak\'ow, Poland}
\date{\today}
\maketitle

\begin{abstract}
Topological and geometrical properties of the set of mixed quantum states in the
$N-$dimensional Hilbert space are analysed. Assuming that the corresponding
classical dynamics takes place on the sphere we use the vector SU(2)
coherent states and the generalised Husimi distributions to define the Monge
distance between two arbitrary density matrices.
The Monge metric has a simple semiclassical interpretation and induces
a non-trivial geometry. Among all pure states the distance from the maximally
mixed state $\rho_*$, proportional to the identity matrix, admits the largest
value for the coherent states, while the delocalized `chaotic' states are close
to $\rho_*$. This contrasts the geometry induced by the standard (trace,
Hilbert-Schmidt or Bures) metrics, for which the distance from $\rho_*$
is the same for all pure states. We discuss possible physical consequences
including unitary time evolution and the process of decoherence.
We introduce also a simplified Monge metric,
 defined in the space of pure
quantum states and more suitable for numerical computation.
\end{abstract}


\vskip 0.4cm

\begin{center}
{\small e-mail: $^1$karol@tatry.if.uj.edu.pl \quad
$^2$slomczyn@im.uj.edu.pl}
\end{center}

\vskip 0.5cm

\vskip 1.0cm

\newpage

\section{Introduction}

Consider two quantum states described by the density matrices $\rho_1$ and
$\rho_2$. What is their distance in the space of quantum states? One should not
expect a unique, canonical answer for this question. On the contrary, several
possible distances can be defined, related to different metrics in this space.
As usual, each solution possesses some advantages and some drawbacks; each might
be useful for different purposes.

Perhaps the simplest possible answer is given by the norm of the difference. The
trace norm leads to the {\sl trace distance}

\begin{equation}
D_{tr}(\rho_1,\rho_2) = \operatorname*{tr} \sqrt{ (\rho_1-\rho_2)^2}
\label{trano}
\end{equation}
used by Hillery \cite{hi87,hi89} to describe the non-classical properties of
quantum states and by Englert \cite{en96} to measure the distinguishability of
mixed states. In a similar way the Frobenius norm results in the
{\sl Hilbert-Schmidt distance}

\begin{equation}
D_{HS}(\rho_1,\rho_2) = \sqrt{\operatorname*{tr} [ (\rho_1-\rho_2)^2]}
\label{HS}
\end{equation}
often used in quantum optics \cite{ko95,wul95,dmmw99}.

Another approach based on the idea of {\sl purification} of a mixed quantum
state leads to the {\sl Bures distance} \cite{bu69,ul76}. An explicit formula
for the Bures distance was found by H{\"{u}}bner \cite{hu92}

\begin{equation}
D_{Bures}(\rho_1,\rho_2)=\sqrt{ 2\bigl(1-{\rm {tr} [(\rho
_1^{1/2}\rho_2\rho_1^{1/2})^{1/2}]\bigr) }} \, \text{,}
\label{Bures}
\end{equation}
and various properties of this distance are a subject of a considerable interest
(see \cite{u94,di95,ps95,ul95,sl96,di99,leru99}). It was
shown by Braunstein and Caves \cite{bc94} that for neighbouring density matrices
the Bures distance is proportional to the statistical distance introduced by
Wootters \cite{wo81} in the context of measurements which optimally resolve
neighbouring quantum states.

Note that for pure states $\rho_1 = |\varphi_1\rangle\langle\varphi_1|$ and 
$\rho_2 = |\varphi_2\rangle\langle\varphi_2|$ we can easily calculate the above 
standard distances, namely

\begin{equation}
D_{tr}(|\varphi_1\rangle, |\varphi_2\rangle) = 2\sqrt{1-p} \, \text{,}
\label{puretr}
\end{equation}

\begin{equation}
D_{HS}(|\varphi_1\rangle, |\varphi_2\rangle) = \sqrt{2\left( 1-p\right) } \,
\text{,}
\label{pureHs}
\end{equation}
and

\begin{equation}
D_{Bures}(|\varphi_1\rangle, |\varphi_2\rangle) =
\sqrt{2\left( 1 - \sqrt{p} \right) } \, \text{,}
\label{pureBures}
\end{equation}
where the 'transition probability' $p = | \langle \varphi_1 | \varphi_2 \rangle
|^2=\cos^2(\Xi/2)$. The angle $\Xi$ equals to the Fubini-Study distance 
$D_{FS}(\varphi_1,\varphi_2)$ in the space of pure states, and for $N=2$ it is 
just the angle between the corresponding points of the Bloch sphere \cite{bh99}. 
The Fubini-Study metric defined by

\begin{equation} 
D_{FS}(\varphi_1,\varphi_2) = {\rm arccos}(2p-1)= 2\ {\rm arccos}\sqrt{p}, 
\label{FS} 
\end{equation} 
corresponds to the geodesic distance in the complex projective space 
(see e.g. \cite{bh99}) and for infinitesimally small values of $p$ becomes 
proportional to any of standard distances.

In a recent paper \cite{zs98} we introduced the {\sl Monge metric} $D_M$ in the
space of density operators belonging to an infinite-dimensional separable
Hilbert space ${\cal {H}}$. The Monge metric fulfils the following
{\sl semiclassical property}: the distance between two harmonic oscillator
(Glauber) coherent states $|\alpha_1\rangle$ and $|\alpha_2\rangle $ localised
at points ${a_1}$ and $a_2$ of the classical phase space $\Omega={\mathbb C}$ is
equal to the Euclidean distance $d$ between these points

\begin{equation}
D_{M}(|\alpha_1\rangle ,|\alpha_2\rangle )=d (a_1,a_2) \, \text{.}
\label{clasprop}
\end{equation}
In the semiclassical regime this condition is rather natural, since the
quasi-probability distribution of a quantum state tends to be strongly localised
in the vicinity of the corresponding classical point. A motivation to study such
a distance stems from the search for quantum Lyapunov exponent, where a link
between distances in the Hilbert space and in the classical phase space is
required \cite{zsw93}. Our construction was based on the Husimi representation
of a quantum state $\rho $ given by \cite{hu40}
\begin{equation}
H_{\rho}(\alpha ):=\frac {1} {\pi} \langle \alpha |\rho |\alpha \rangle \,
\text{,}
\label{hus1}
\end{equation}
for $\alpha \in \mathbb C$. The Monge distance $D_M$ between two arbitrary
quantum states was defined as the Monge-Kantorovich distance between the
corresponding Husimi distributions \cite{zs98}.

Although the Monge-Kantorovich distance is not easy to calculate
for two or more dimensional problems, it satisfies
the semiclassical property (\ref{clasprop}),
crucial in our approach.   On the other hand, one
could not use for this purposes any
`simpler' distances between the Husimi distributions, like e.g.
   $L_1$ or $L_2$ metrics, because the
semiclassical property does not hold in these cases.
 Moreover, this property is not
  fulfilled for any of the standard distances
in the space of density matrices (trace,
  Hilbert-Schmidt or Bures distances).
Consider two arbitrary pure quantum states $|\varphi_1\rangle$ and
$|\varphi_2\rangle \in {\cal H}$
and the corresponding density operators \linebreak $\rho_1 \noindent =
|\varphi_1\rangle
\langle\varphi_1|$ and $\rho_2=|\varphi_2\rangle\langle\varphi_2|$. If the
states are orthogonal, the standard distances between them do not depend on
their localisation in the phase space. For example the Hilbert--Schmidt and the
Bures distances between two different Fock states $|n\rangle$ and $|m\rangle$
are equal to $\sqrt{2}$, and the trace distance is equal to $2$. Although the
state $|1\rangle$ is localised in the
phase space much closer to the state $|2\rangle$ then to $|100\rangle$, this
fact is not reflected by any of the standard distances.
 Clearly, the same concerns a nonlinear
  function of the Hilbert-Schmidt distance, which satisfy the
semiclassical condition (\ref{clasprop}) and was recently introduced in
\cite{dmmw00}. On the other hand, the
 Monge distance is capable to reveal the
phase space structure of the quantum states, since
$D_M(|m\rangle,|n\rangle) =\left| a_{m} - a_{n}\right| $, where $a_k =
\sqrt{\pi} \binom{2k}{k}(2k+1)/2^{2k+1}
\sim \sqrt{k}$ \thinspace (see \cite{zs98}).

In this paper we propose an analogous construction for a classical compact phase
space and the corresponding finite-dimensional Hilbert spaces ${\cal H}_N$. In
particular we discuss the $N=(2j+1)-$dimensional Hilbert spaces generated by the
angular momentum operator $J$. In the classical limit the quantum number $j$
tends to infinity and the classical dynamics takes place on the sphere $S^2$.
However, the name {\sl Monge metric on the sphere} should not be interpreted
verbally: the metric is defined in the space of density matrices, while the
connection with the sphere is obtained via the $SU(2)$ vector coherent
states, used in the construction to
 represent a quantum state by its generalised
Husimi distribution. In general,
 the Monge distance in the space of quantum
states can be defined with respect to an arbitrary classical phase space
$\Omega$.

This paper is organised as follows. In Sect.~II we review some properties of
pure and mixed quantum states in a finite-dimensional Hilbert space. In
Sect.~III we recall the definition of the Monge metric based on the Glauber
coherent states and extend this construction to an arbitrary set of
(generalised) coherent states. We analyse basic properties of such defined
metric and its relation to other distances in the space of density operators.
The case where the classical phase space is isomorphic with the sphere $S^2$,
corresponding to the $SU(2)$ coherent states, is considered in Sect.~IV. We
compute the Monge distance between certain pure and mixed states, and compare
the results with other distances (trace, Hilbert-Schmidt, and Bures). In
particular, we give the formulae for the Monge distance between two coherent
states (for arbitrary $j$) and between two arbitrary mixed states for $j=1/2$.
In the latter case the geometry induced by the Monge distance coincides with the
standard geometry of the Bloch sphere induced by the Hilbert-Schmidt (or the
trace) distance. However, in the higher dimensions both geometries differ
considerably. Potential physical consequences of our approach are discussed in
Sect.~V. In Sect.~VI we introduce a simplified version of the Monge metric,
defined only in the space of pure quantum states, but better suited for
numerical computation. Finally, some concluding remarks are provided in
Sect.~VII.

\section{Space of mixed quantum states}

\subsection{Topological properties}

Let us consider a {\sl pure quantum state} $|\psi\rangle$ belonging to an
$N-$dimensional Hilbert space ${\cal {H}}_N$. It may be described by a
normalised vector in ${\cal {H}}_N$, or by the density matrix
$\rho_{\psi}=|\psi\rangle\langle\psi|$. Such a state fulfills the purity
condition: $\rho^2_{\phi}=\rho_{\phi}$. The manifold ${\cal P}$, containing all
pure states, is homeomorphic with the complex projective space ${\mathbb C}P^{N-
1}$. This space is $2(N-1)-$dimensional. In the simplest case $N=2$, the
two-dimensional space ${\mathbb C}P^1$ corresponds to the {\sl Bloch sphere}.

To generalise the notion of pure states one introduces the concept of {\sl mixed
quantum states}. They are represented by $N\times N$ positive Hermitian matrices
$\rho $, which satisfy the trace condition $\operatorname*{tr}\rho =1$. Any density matrix may
be diagonalized and represented by

\begin{equation}
\rho =VEV^{\dagger } \, \text{,}
\label{diag}
\end{equation}
where $V$ is unitary, while a diagonal matrix of eigenvalues $E$ contains only
non-negative entries: $E_{i}\geq 0; \thinspace i=1,\dots ,N$. For each pure
state all entries of $E$ are equal to zero, but one equal to unity. Due to the
trace condition $\sum_{i=1}^{N}E_{i}=1$. It means that the set of all such
matrices $E$ forms an $(N-1)-$dimensional simplex ${\cal S}_N$ in ${\mathbb
R}^N$. Let $B$ be a diagonal unitary matrix. Since
\begin{equation}
\rho =VEV^{\dagger }=VBEB^{\dagger }V^{\dagger } \, \text{,}
\label{diag2}
\end{equation}
therefore the matrix $V$ is determined up to $N$ arbitrary phases entering $B$.
On the other hand, the matrix $E$ is defined up to a permutation of its entries.
The form of the set of such permutations depends on the character of the
degeneracy of the spectrum of $\rho$.

\begin{figure}
 \hspace*{4.3cm}
 \epsfxsize=11.0cm
\epsfbox{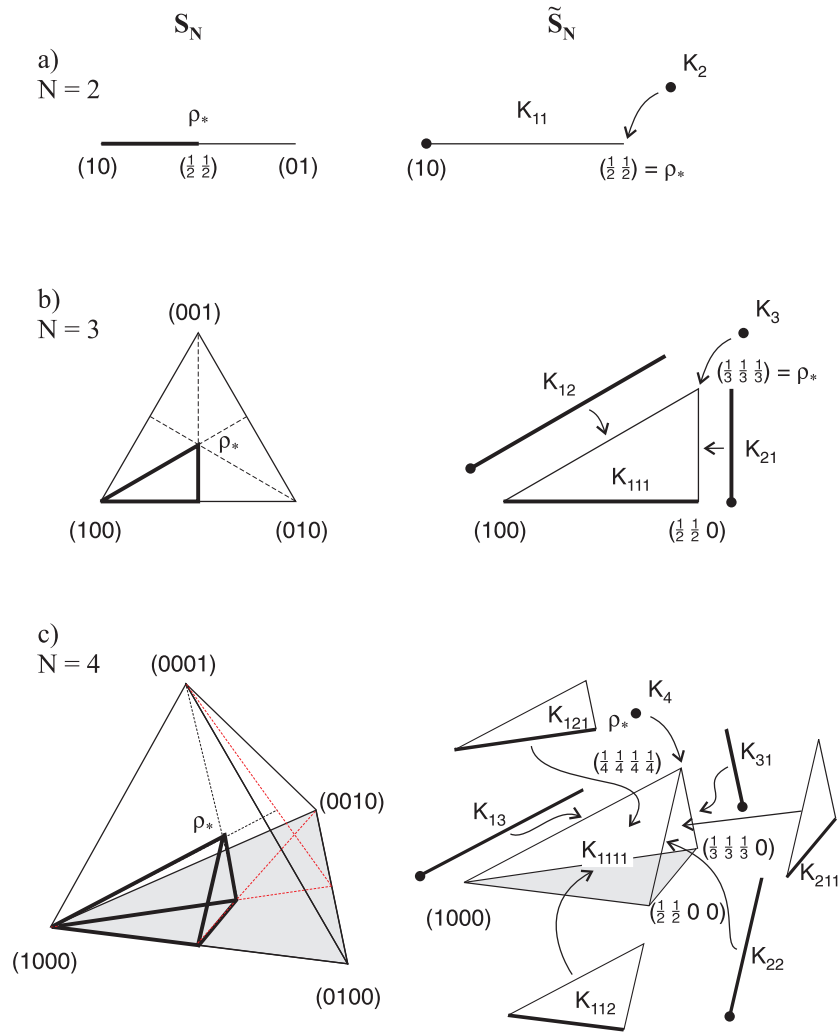} \\
\caption{$(N-1)-$dimensional simplex ${\cal S}_N$ of diagonal density
matrices of size $N$ and its antisymmetric part ${\widetilde{\cal S}}_N$
for a) $N=2$, b) $N=3$, and c) $N=4$. The simplex ${\widetilde{\cal S}}_N$,
enlarged at the right hand side, can be decomposed into $2^N$ parts.
$N$ numbers in brackets denote coordinates in the original $N-$dimensional space
of eigenvalues. Corners of ${\widetilde{\cal S}}_N$ represent pure states
(density matrices of rank one), edges - matrices of rank two, faces - matrices
of rank two. Bold lines (grey faces) symbolise boundary of ${\widetilde{\cal
S}}_N$.}
\label{f1}
\end{figure}

Representation (\ref{diag2}) makes the description of some 
topological properties of the $(N^2-1)-$dimensional space ${\cal M}$ easier 
\cite{ach93,bbms98}. We introduce the following notation. We write $T^N 
= (S_1)^N = [U(1)]^N$ for the $N-$dimensional torus. Identifying points of 
${\cal S}_N$ which have the same coordinates (but ordered in a different way) we 
get an asymmetric simplex $\widetilde{\cal S}_{N}$. Equivalently, one can divide 
${\cal S}_N$ into $N!$ identical simplexes and take any of them. The asymmetric 
simplex $\widetilde{\cal S}_{N}$ can be decomposed in the following natural way 
 
\begin{equation} 
\widetilde{\cal S}_{N} = \bigcup_{k_1+\dots+k_n=N} K_{k_1,\dots,k_n} \, \text{,} 
\label{simplex} 
\end{equation} 
where $n=1,\dots,N$ denotes the number of different coordinates of a given point 
of $\widetilde{\cal S}_{N}$, $k_1$ the number of occurrences of the largest 
coordinate, $k_2$ the second largest, etc. Observe that $K_{k_1,\dots,k_n}$ is 
homeomorphic with the set $G_n$, where $G_1$ is a single point, $G_2$ a 
half-closed interval, $G_3$ an open triangle with one edge but without corners, 
and generally, $G_n$ is an $(n-1)-$dimensional simplex with one $(n-2)-$dimensional hyperface without boundary (the latter is homeomorphic with an 
$(n-2)-$dimensional open simplex). There are $N$ ordered eigenvalues: $E_1\ge 
E_2\ge \cdots \ge E_N$, and $N-1$ independent relation operators 
`larger or equal', which makes all together $2^{N-1}$ different 
possibilities. Thus, $\widetilde{\cal S}_{N}$ consists of $2^{N-1}$ 
parts, out of which $\binom{N-1}{m-1}$ parts are homeomorphic with 
$G_m$, when $m$ ranges from $1$ to $N$. The decomposition of the asymmetric 
simplex $\widetilde{\cal S}_{N}$ is illustrated in Fig.~1 for the simplest cases 
$N=2,3$, and $4$. 

Let us denote the part of the space ${\cal M}$ related to the spectrum in
$K_{k_1,\dots,k_n}$ ($n$ different eigenvalues, the largest eigenvalue has $k_1$
multiplicity, the second largest $k_2$ etc.) by ${\cal M}_{k_1,\dots,k_n}$.
A mixed state $\rho$ with this kind of the spectrum remains invariant under
arbitrary unitary rotations performed in each of the $k_i-$dimensional subspaces
of degeneracy. Therefore the unitary matrix $B$ has a block diagonal structure
with $n$ blocks of size equal to $k_1,\dots,k_n$ and

\begin{equation}
{\cal M}_{k_1,\dots,k_n}\sim \lbrack U(N)/(U(k_1) \times \cdots
\times U(k_n)) \rbrack \times G_{n} \, \text{,}
\label{repr2}
\end{equation}
where $k_1+\dots+k_n=N$ and $k_i>0$ for $i=1,\dots,n$.
Thus $\cal M$ has the structure
\begin{equation}
{\cal M} \sim \bigcup_{k_1+\dots+k_n=N} {\cal M}_{k_1,\dots,k_n}\sim
\bigcup_{k_1+\dots+k_n=N}
\lbrack U(N)/(U(k_1) \times \cdots \times U(k_n))
\rbrack \times G_{n} \, \text{,}
\label{repr3}
\end{equation}
where the sum ranges over all partitions of $N$.
The group of rotation matrices $B$ equivalent to
$\Gamma = U(k_1)\times U(k_2)\times \cdots \times U(k_n)$
is called the {\sl stability group} of $U(N)$.

For $N=2$ we have ${\cal M}_{1,1} \sim [U(2)/T^2] \times G_{2} \sim S^2 \times
G_{2}$ and ${\cal M}_{2}\sim \{\rho_*\}$, so the
space ${\cal M}$ has the topology of a two-dimensional ball - the Bloch sphere
and its interior. This case and also cases $N=3,4$ are analysed in detail in
Tab.~1.

\bigskip

\hskip -0.3cm
\begin{tabular}
[c]{|c|c|c|c|c|c|c|}\hline
$N$ & Label & Decomposition & Subspace
& \parbox{4cm}{\centering Part of the \\ asymmetric simplex} &
\ \ Topological Structure & \parbox{2cm}{\centering Dimension \\ $D=D_1+D_2$}
\\\hline\hline
$1$ & ${\cal M}_{1}$ & $1$ & $E_{1}$ & point & $[U(1)/U(1)]\times
G_{1}=\{\rho_{\ast}\}$ & $0=0+0$
\\\hline\hline
$2$ & ${\cal M}_{11}$ & $1+1$ & $E_{1}>E_{2}$ & line with left edge &
$[U(2)/T^{2}]\times G_{2}$ & $ 3=2+1$
\\\cline{2-7} & ${\cal M}_{2}$ & $2$ & $E_{1}=E_{2}$
& right edge & $[U(2)/U(2)]\times
G_{1} =\{\rho_{\ast}\}$ & $0=0+0$
\\\hline\hline
& ${\cal M}_{111}$ & $1+1+1$ & $E_{1}>E_{2}>E_{3}$
& \parbox{4cm}{\centering triangle with base \\ without corners} &
$[U(3)/T^{3}]\times G_{3}$ & $8=6+2$
\\\cline{2-7}
$3$ & ${\cal M}_{12}$ & $1+2$ & $E_{1}>E_{2}=E_{3}$ & edges with &
$[U(3)/(U(2)\times T)]\times G_{2}$ & $5=4+1$
\\\cline{2-4}
& ${\cal M}_{21}$ & $2+1$ & $E_{1}=E_{2}>E_{3}$ & lower corners & &
\\\cline{2-7}
& ${\cal M}_{3}$ & $3$ & $E_{1}=E_{2}=E_{3}$ & upper corner &
$[U(3)/U(3)]\times G_{1}=\{\rho_{\ast}\}$ & $0=0+0$
\\\hline\hline
& ${\cal M}_{1111}$ & $1+1+1+1$ & $E_{1}>E_{2}>E_{3}>E_{4}$ &
\parbox{4cm}{\centering interior of tetrahedron with bottom face} &
$[U(4)/T^{4}]\times G_{4}$ & $15=12+3$
\\\cline{2-7}
& ${\cal M}_{112}$ & $1+1+2$ & $E_{1}>E_{2}>E_{3}=E_{4}$ & & &
\\\cline{2-4}
& ${\cal M}_{121}$ & $1+2+1$ & $E_{1}>E_{2}=E_{3}>E_{4}$ & faces without side
edges & $[U(4)/(U(2)\times T^{2})]\times G_{3}$ & $12=10+2$
\\\cline{2-4}
$4$ & ${\cal M}_{211}$ & $2+1+1$ & $E_{1}=E_{2}>E_{3}>E_{4}$ & & &
\\\cline{2-7}
& ${\cal M}_{13}$ & $1+3$ & $E_{1}>E_{2}=E_{3}=E_{4}$ & &
$[U(4)/(U(3)\times T)]\times G_{2}$ & $7=6+1$
\\\cline{2-4}
& ${\cal M}_{31}$ & $3+1$ & $E_{1}=E_{2}=E_{3}>E_{4}$ &
edges with lower corners & &
\\\cline{2-4}
\cline{6-7}
& ${\cal M}_{22}$ & $2+2$ & $E_{1}=E_{2}>E_{3}=E_{4}$ & & $[U(4)/(U(2)\times
U(2))]\times G_{2}$ & $9=8+1$
\\\cline{2-7}
& ${\cal M}_{4}$ & $4$ & $E_{1}=E_{2}=E_{3}=E_{4}$ & upper corner &
$[U(4)/U(4)]\times G_{1}=\{\rho_{\ast}\}$ & $0=0+0$
\\\hline
\end{tabular}

\medskip

Table 1. \ Topological structure of the space of mixed quantum states for a
fixed number of levels $N$. The group of unitary matrices of size $N$ is denoted
by $U(N)$, the unit circle (one-dimensional torus $\sim U(1)$) by $T$, while
$G_{n}$ stands for a part of an $(n-1)-$dimensional asymmetric simplex
defined in the text. Dimension $D$ of the component ${\cal M}_{k_1,\dots,k_n}$
equals $D_1+D_2$, where $D_1$ denotes the dimension of the quotient space
$U(N)/\Gamma$, while $D_2=n-1$ is the dimension of the part of the eigenvalues
simplex homeomorphic with $G_n$.
\bigskip

Note that the part $M_{1,\dots,1}$ represents generic, non-degenerate spectrum.
In this case all elements of the spectrum of $\rho$ are different and
the stability group $H$ is equivalent to an $N$-torus
\begin{equation}
{\cal M}_{1,\ldots ,1} \sim [U(N)/T^N] \times G_N \, \text{.}
\label{repr1}
\end{equation}

Above representation of generic states enables us to define a product measure in
the space ${\cal M}$ of mixed quantum states. For this end, one can take the
uniform (Haar) measure on $U(N)$ and a certain measure on the simplex
${\cal S}_N$ \cite{zhsl98,sl98}. The coordinates of a point on the simplex may
be generated \cite{zy99} by squared moduli of components of a random orthogonal
(unitary) matrix \cite{pzk98}.

The other $2^{N-1}-1$ parts of $\cal M$ represent various kinds of
degeneracy and have measure zero. The number of non-homeomorphic parts is equal
to the number $P(N)$ of different representations of the number $N$ as the sum
of positive natural numbers. Thus $P(N)$ gives the number of different
topological structures present in the space $\cal M$. For $N=1,2,\dots,10$ the
number $P(N)$ is equal to $1,2,3,5,7,11,15,22,30$ and $42$, while for larger $N$
there is described by the asymptotic Hardy-Ramanujan formula \cite{hr18},
$P(N) \sim \exp \left( \pi \sqrt{2N/3}\right)/4\sqrt{3}N$.

In the extreme case of $N$-fold degeneracy, $E_{i} \equiv 1/N$, the subspace
${\cal M}_{N}\sim \lbrack U(N)/(U(N)\times T^{0}) \rbrack
\times G_{1}\sim G_{1}$, so it degenerates to a single point. This distinguishes
the maximally mixed state $\rho _{\ast}:= {\bf I}/N$, which will play a crucial
role in subsequent considerations. For the manifold of pure states $n=2$ and
$k_1 = 1, k_2 = N-1$ (since $E_{1}=1$, $E_{2}=\cdots =E_{N}=0$) and so
${\cal P}\sim \lbrack U(N)/(U(N-1)\times U(1)) \rbrack \times (1,0,\dots,0)
\sim {\mathbb C}P^{N-1}$.
In the case $N=2$ it can be identified with the Bloch sphere $S^2$.

On the other hand, it is well known that $\cal M$ itself has a structure of a
simplex with the boundary contained in the hypersurface det$\rho = 0$,
with rank 1 matrices (pure states - $\cal P$) as `corners', rank 2 as `edges',
etc., and with the point $\rho_{\ast}$ `in the middle' (see \cite{cl93,bcl95}
for a formal statement and \cite{be98} for a nice intuitive discussion).

Let us mention in passing that the quotient space
appearing in (\ref{repr2})
\begin{equation}
{\mathbb F}:= {U(N) \over U(k_1) \times U(k_2)\times
\cdots\times U(k_n)}
\label{flag}
\end{equation}
is called a {\sl flag manifold}, and in a special case
\begin{equation}
Gr(k,N):= {U(N) \over U(k) \times U(N-k)}
\label{grass}
\end{equation}
a {\sl Grassman manifold}. For a fuller discussion of the topological structure
of $\cal M$ (especially for $N=4$) we refer the reader to \cite{ach93}.

\subsection{Metric properties}

The density matrix of a pure state $\rho_{\psi}=|\psi\rangle\langle\psi|$ may be
represented in a suitable basis by a matrix with the first element equal
to unity and all others equal to zero. Due to this simple form it is
straightforward to compute the standard distances between $\rho_{\psi}$ and
$\rho_*$ directly from the definitions recalled in Sect.~1. Results do depend on
the dimension $N$, but are independent on the pure state $|\psi\rangle$, namely

\begin{equation}
D_{tr}(\rho_{\psi},\rho_*) = 2- \frac{2}{N} \, , \qquad
D_{HS}(\rho_{\psi},\rho_*) = \sqrt{ 1 - \frac{1}{N}} \, , \qquad
D_{Bures}(\rho_{\psi},\rho_*) = \sqrt{2- \frac{2}{\sqrt{N}}} \, \text{.}
\label{pure}
\end{equation}
In the sense of the trace, the Hilbert-Schmidt, or the Bures metric the
$2(N-1)-$dimensional space of pure states ${\cal P}$ may be therefore
considered as a part of the $(N^2-1)-$dimensional sphere centred at $\rho_*$ of
the radius $r$ depending on $N$ and on the metric used.
From the point of view of these standard metrics,
no pure state on ${\cal P}$ is distinguished; all of them
are equivalent. It is easy to show that the distance of any mixed state from
$\rho_*$ is smaller than $r$, in the sense of each of the standard metrics. Thus
the space of mixed states ${\cal M}$ lays inside the sphere $S^{N^2-2}$
embedded in ${\mathbb R}^{N^2-1}$, although, as discussed above, its topology
(for $N>2$) is much more complicated than the topology of the
$(N^2-1)-$dimensional disk.

The degree of mixture of any state may be measured, e.g., by the {\sl von
Neumann entropy} $S=- {\rm Tr}\rho\ln\rho=-\sum_{i=1}^N E_i\ln(E_i)$. It
varies
from zero (pure states) to ln($N$) (the maximally mixed state $\rho_*$). Let us
briefly discuss a simple kicked dynamics, generated by a Hamiltonian represented
by a Hermitian matrix $H$ of size $N$. It maps a state $\rho$
into
\begin{equation}
\rho^{\prime}= e^{iH} \rho e^{-iH} \, \text{,}
\label{dynamics}
\end{equation}
where the kicking period is set to unity.

Such a unitary quantum map does not change the eigenvalues of $\rho $, so the
von Neumann entropy is conserved. In particular, any pure state is mapped by
(\ref{dynamics}) into a pure state. Any mixed state $\rho $, which commutes with
$H$, is not affected by this dynamics. Assume the Hamiltionian $H$ to be
generic, in the sense that its $N$ eigenvalues are different. Then its invariant
states form an $(N-1)-$dimensional subspace ${\cal{I}}_H \subset M$,
topologically equivalent to ${\cal S}_N$. In the generic case of non-degenerate
Hamiltonian it contains only $N$ pure states: the
eigenstates of $H$. Note that the invariant subspace ${\cal I}_H$ always
contains $\rho _{\ast}$.

Moreover, the standard distances between two states are conserved under the
action of a unitary dynamics, i.e.

\begin{equation}
D_{s}(\rho _{1},\rho _{2})=D_{s}(\rho _{1}^{\prime },\rho _{2}^{\prime }) \,
\text{,}
\label{invar}
\end{equation}
where $D_{s}$ denotes one of the distances: $D_{tr}$, $D_{HS}$
or $D_{Bures}$. Therefore, the unitary dynamics given by (\ref{dynamics}) can be
considered as a generalised rotation in the $(N^{2}-1)-$dimensional space ${\cal
M}$, around the $(N-1)-$dimensional `hyperaxis' ${\cal I}_H$, which is
topologically equivalent to the simplex $S^{N-1}$. In the simplest
case, $N=2$, it is just a standard rotation of the Bloch ball around the axis
determined by $H$. For example, if $H=\alpha J_{z}$, where $J_z$ is the third
component of the angular momentum operator $J$, it is just the rotation by
angle $\alpha $ around the $z$ axis joining both poles of the Bloch sphere. The
set ${\cal I}_H$ of states invariant with respect to this
dynamics consists of all states diagonal in the basis of $J_{z}$: the mixed
states with diag$(\rho)=\{a,1-a\}$ ($a\in (0,1)$) and two pure states,
$|1/2,1/2\rangle $ for $a=1$, and $|1/2,-1/2\rangle $ for $a=0$.

\section{Monge distance between quantum states}

\subsection{Monge transport problem and the Monge-Kantorovich distance}

The original Monge problem, formulated in 1781 \cite{mo1781}, emerged from
studying the most efficient way of transporting soil \cite{ra84}:

{\sl Split two equally large volumes into infinitely small particles and then
associate them with each other so that the sum of products of these paths of the
particles over the volume is least. Along which paths must the particles be
transported and what is the smallest transportation cost?}

Consider two probability densities $Q_1$ and $Q_2$ defined in an open set
$\Omega \subset R^n$, i.e., $Q_i \geq 0$ and $\int_\Omega Q_i(x) d^n x = 1$ for
$i=1,2$. Let $V_1$ and $V_2$, determined by $Q_i$, describe the initial and the
final location of `soil': $V_i=\{\left( x,y\right) \in \Omega \times R^{+}:0
\leq y \leq Q_i(x)\}$. The integral $\int_{V_i} d^nx\,dy$ \/is equal to the
unity due to normalisation of $Q_i$. Consider $C^1$ one-to-one maps $T:\Omega\to
\Omega$ which generate volume preserving transformations $V_1$ into $V_2$, i.e.,

\begin{equation}
Q_1\left( x\right) =Q_2\left( Tx\right) \left| T^{\prime}(x)\right|
\label{trans}
\end{equation}
for all $x\in \Omega$, where $T^{\prime}(x)$ denotes the Jacobian of the map $T$
at point $x$. We shall look for a transformation giving the minimal displacement
integral and define the {\sl Monge distance} \cite{ra84,ra91}

\begin{equation}
D_M(Q_1,Q_2):={\inf}\int_{\Omega}|x-T(x)|Q_1\left( x\right) d^n x \, \text{,}
\label{monge}
\end{equation}
where the infimum is taken over all $T$ as above. If the optimal transformation
$T_M$ exists, it is called a {\sl Monge plan}. Note that in this formulation of
the
problem the `vertical' component of the soil movement is neglected. The problem
of existence of such a transformation was solved by Sudakov \cite{su76}, who
proved that a Monge plan exists for $Q_1,Q_2$ smooth enough (see also \cite
{eg99}). The above definition can be extended to an arbitrary metric space
$(\Omega,d)$ endowed with a Borel measure $m$. In this case one should put
$d(x,T(x))$ instead of $|x-T(x)|$ and $dm(x)$ instead of $d^{n}x$ in formula
(\ref{monge}), and take the infimum over all one-to-one and continuous $T:\Omega
\to \Omega$ fulfilling $\int_A Q_1 dm = \int_{T^{-1}(A)} Q_2 dm$ for each Borel
set $A \subset \Omega$. In fact we can also measure the Monge distance between
arbitrary two probability measures in a metric space $(\Omega,d)$. For
$\mu,\nu$ - probability measures on $(\Omega,d)$ we put
\begin{equation}
D_M(\mu,\nu):={\inf}\int_{\Omega} d(x,T(x)) d\mu(x) \ \text{,}
\label{mongearbmes}
\end{equation}
where the infimum is taken over all one-to-one and continuous $T: \Omega \to
\Omega$ such that $\mu(A) = \nu(T^{-1}(A))$ for each Borel set $A \subset
\Omega$. To avoid the problem of the existence of a Monge plan Kantorovich
\cite{ka42,ka48} introduced in 40s the `weak' version of the original Monge's
mass allocation problem and proved his famous variational principle (see
Proposition~1). For this and other interesting generalisations of the Monge
problem consult the monographs by Rachev and R{\"u}schendorf \cite{ra91,rr98}.

In some cases one can find the Monge distance analytically.
For the one-dimensional case, $\Omega = \mathbb R$, the Monge distance can be
expressed
explicitly with the help of distribution functions
$F_i(x)=\int_{-\infty }^x Q_i(t)dt$, $i=1,2$. Salvemini
obtained the following solution of the problem \cite{sa43}
\begin{equation}
D_M(Q_1,Q_2)=\int_{-\infty }^{+\infty }|F_1(x)-F_2(x)|dx \, \text{.}
\label{salv}
\end{equation}

Several two-dimensional problems with some kind of symmetry can be reduced to
one-dimensional problems, solved by (\ref{salv}). In the general case one can
estimate the Monge distance numerically \cite{wzs98}, relying on algorithms of
solving the transport problem, often discussed in handbooks of linear
programming \cite{wc81}.

\begin{figure} 
 \begin{center} 
\ 
 \vskip -1.8cm 
\includegraphics[width=14cm,angle=270]{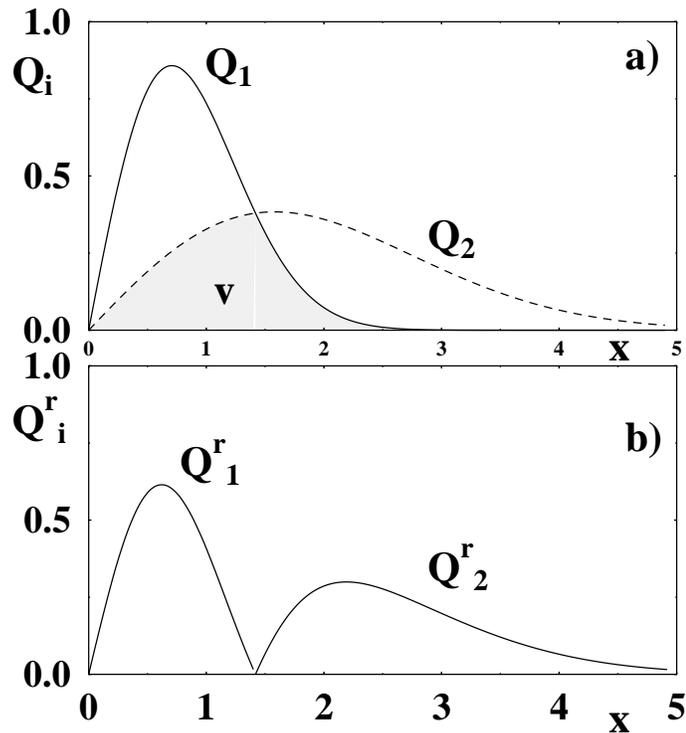} 
 \end{center} 
\vskip -2.5cm 
\caption{Estimation of the Monge distance between two 
overlapping distributions $Q_1$ and $Q_2$ (a) by the distance 
between the reduced distributions ${Q}^r_i(x) = Q_i(x)-V(x)$ (b), 
where $V(x)$ denotes the overlap.} 
\label{f2} 
\end{figure} 

According to definition (\ref{monge}) taking an arbitrary map $T$ which fulfils 
(\ref{trans}) we obtain an {\sl upper bound} for the Monge distance $D_M$. 
Another two methods of estimating the Monge distance are valid for a compact 
metric space $(\Omega,d)$ equipped with a finite measure $m$. The first method 
may be used to obtain lower bounds for $D_M$. It is based on the proposition 
proved by Kantorovich in 40s \cite{ka42,ka48} (see also \cite{ra91,rr98,eg99}). 
 
\medskip 
 
{\bf Proposition 1.} ({\sl variational formula for the Monge-Kantorovich 
metric}) 
\begin{equation} 
D_M(Q_1,Q_2) = \max \left| \int_{\Omega} f(x)(Q_1(x)- Q_2(x))dm(x) \thinspace 
\right| \, \text{,} 
\label{varKan} 
\end{equation} 
where the supremum is taken over all $f$ fulfilling the condition $|f(x)-f(y)| 
\leq d(x,y)$ for all $x,y \in \Omega$ (weak contractions). 
 
To obtain another {\sl upper bound} for $D_M$ we may apply the following 
simple estimate: 
 
\medskip 
 
{\bf Proposition 2.} 
\begin{equation} 
D_M(Q_1,Q_2) \leq (\Delta/2) D_{L_1}(Q_1,Q_2), 
\label{ineq1} 
\end{equation} 
where $D_{L_1}(Q_1,Q_2) = \int_{\Omega}|Q_1(x)-Q_2(x)| dm(x)$ and 
$\Delta=\operatorname*{diam}(\Omega)$.

The intuitive explanation of this fact is the following. Let $v$ be the volume 
of the `overlap' of the probability distributions $Q_1$ and $Q_2$, i.e., 
$v = \int_{\Omega} V(x) dm(x)$, where $V = \min\{Q_1,Q_2\}$. 
Then $D_M(Q_1,Q_2) \leq (1-v) \Delta$, because the number $(1-v)$ represents the 
part of the distribution to be moved and the largest possible classical 
distance on $\Omega$ is smaller than or equal to $\Delta$. Moreover, 
$D_{L_1}(Q_1,Q_2) = 2(1-v)$, which proves the assertion. Although Fig.~2 
presents the corresponding picture for the simplest, one-dimensional case, 
Proposition~2 is valid for an arbitrary metric space. For the formal proof see 
Appendix~A. 

\subsection{The Monge distance - harmonic oscillator coherent states}

In \cite{zs98} we defined a {\sl `classical'} distance between two quantum
states $\rho_1$ and $\rho_2$ via the Monge distance between the corresponding
Husimi distributions $ H_{\rho_1}$ and $ H_{\rho_2}$:
\begin{equation}
D_{M}(\rho_1,\rho_2):= D_M \left({H_{\rho_1},H_{\rho_2}}\right) \, \text{,}
\label{mondef}
\end{equation}
where $H_{\rho_{i}}$ are given by formula (\ref {hus1}).
Observe that the family of harmonic oscillator coherent states
$|\alpha\rangle$, parameterised by a complex number $\alpha$, is
implicitly present in this definition.

The Monge distance satisfies the {\sl semiclassical property}: the distance
between any two Glauber coherent states, represented by Gaussian Husimi
distributions localised at points $a_1$ and $a_2$, is equal to the classical
distance $|a_1-a_2|$ in the complex plane \cite{zs98}.

\subsection{The Monge distance - general case and basic properties}

The above construction, originally performed for the complex plane with the help
of the harmonic oscillator coherent states, may be extended to arbitrary
generalised coherent states of Perelomov \cite{pe86} defined on a compact
classical phase space. Let $G$ be a compact Lie group, $G \ni g \to R_g \in
{\cal H}$ its irreducible unitary representation in the Hilbert space $\cal
H$, and $\Upsilon $ the subgroup of $G$, which consists of all elements $y
\in G$ leaving the reference state $|\kappa \rangle \in {\cal P}$ invariant
(i.e. $R_y |\kappa \rangle \sim |\kappa \rangle$). Define $\Omega = G/
\Upsilon$ and $|\eta \rangle = R_{\eta}|\kappa
\rangle$ for $[\eta] \in G/ \Upsilon$. Note that $| 1 \rangle = | \kappa
\rangle$, where $1$ is the group $G$ unit.
Consider a family of the {\sl generalised
coherent states} $\Omega \ni \eta \to |\eta \rangle \in {\cal P}$. It satisfies
the identity resolution $\int_{\Omega} |\eta \rangle \langle \eta| dm(\eta) =
{\bf I}/\dim \cal{H}$, where
$m$ is the natural (translation invariant) measure on the Riemannian manifold
$(\Omega,d)$ normalised by the condition $m(\Omega)=1$, and $d$ is the
Riemannian metric on $\Omega$. Let us denote by ${\cal C}$ the manifold of all
quantum coherent states, isomorphic to $\Omega$, and embedded in the space of
all pure states ${\cal P}$. Note that $\langle \eta | \eta \rangle \equiv 1$.

For the $SU(k)$ coherent states the space $\Omega \sim {\cal C}$ is isomorphic
to ${\mathbb C}P^{k-1}$ and $m$ is the natural Riemannian measure on $\Omega$.
Obviously, the dimension of the Hilbert space ${\cal H}_N$ carrying the
representation of the group equals $N \ge k$, and if $N=k$ all pure states are
$SU(k)$ coherent, and ${\cal C} = {\cal P}$.
For example, in the case of $SU(2)$
vector coherent states the corresponding
classical phase space is the sphere
$S^2 \simeq {\mathbb C}P^1$ \cite{pe86}.
In the simplest case $N=2$ (or $j=1/2$)
pure states are located at the Bloch sphere and are coherent.

Any quantum state $\rho \in \cal{M}$ may be represented by a generalised Husimi
distribution $H_{\rho} : \Omega \to \mathbb{R}^+$ defined by
\begin{equation}
H_{\rho}(\eta ) := N \cdot \langle \eta |\rho |\eta \rangle \, \text{,}
\label{hus2}
\end{equation}
for $\eta \in \Omega$, which satisfies
\begin{equation}
\int_{\Omega} H_{\rho}(\eta ) dm(\eta) = 1 \, \text{.}
\label{inthus}
\end{equation}
In particular, for a pure state $\rho = |\vartheta \rangle \langle \vartheta|$
($|\vartheta \rangle \in \cal{P}$) and $\eta \in \Omega$ we have
\begin{equation}
H_{|\vartheta \rangle \langle \vartheta|}(\eta ) :=
N {|\langle \vartheta | \eta \rangle |}^2 \, \text{.}
\label{huscoh}
\end{equation}

In the sequel we shall assume that for coherent states $|\vartheta \rangle \in
\cal{C}$ ($\vartheta \in \Omega$) the densities
$H_{|\vartheta \rangle \langle \vartheta|}$ tend weakly to the
Dirac-delta measure $\delta_{\vartheta}$ in the {\sl semiclassical limit}, i.e.,
when the dimension of the Hilbert space carrying the representation tends to
infinity.

The Monge distance for the Hilbert space ${\cal H}_N$, the classical phase space
$(\Omega,d)$ and the corresponding family of generalised coherent states
$|\eta\rangle$ is then defined by solving the Monge problem in $\Omega$, in the
full analogy with (\ref{mondef}):

\begin{equation}
D_{M}(\rho_1,\rho_2):= D_M \left({H_{\rho_1},H_{\rho_2}}\right) \, \text{.}
\label{Monge-def}
\end{equation}

The distance $d(\eta,T(\eta))$ between the initial point $\eta$
and its image $T(\eta)$ with respect to the Monge plan has to be computed along
the geodesic lines on the Riemannian manifold $\Omega$.

For compact spaces $\Omega$ the semiclassical condition (\ref{clasprop}) for the
distance between two coherent states becomes weaker:

\medskip

{\bf Property A.} ({\sl semiclassical condition}) Let $\eta_1, \eta_2 \in
\Omega$. Then

\begin{equation}
D_M(|\eta_1\rangle ,|\eta_2\rangle ) \le d (\eta_1,\eta_2) \, \text{,}
\label{clasprop2}
\end{equation}

and

\begin{equation}
D_M(|\eta_1\rangle ,|\eta_2\rangle ) \to d (\eta_1,\eta_2) \qquad \text{(in the
semiclassical limit)} \text{,}
\label{clasprop4}
\end{equation}

where $d$ represents the Riemannian distance between two points in
$\Omega$.

\medskip

To demonstrate (\ref{clasprop2}) it suffices to take for the transformation $T$
in (\ref{monge}) the group translation ${\eta_2}*{\eta_1}^{-1}$ (e.g. the
respective rotation of the sphere $S^2$ in the case of $SU(2)$ coherent states).
However, this transformation needs not to give the
optimal Monge plan. As we shall show in the following section, this is so
for the sphere and the $SU(2)$ coherent states. On the other hand, in the
semiclassical limit (for $SU(2)$ coherent states: $j\to \infty$), the inequality
in (\ref{clasprop2}) converts into the equality, in the full analogy to the
property (\ref{clasprop}), valid for the complex plane and the harmonic
oscillator coherent states. This follows from the fact that the Monge-
Kantorovich metric generates the weak topology in the space of all probability
measures on $\Omega$, and the densities $H_{|\eta_{i} \rangle \langle
\eta_{i}|}$ tend weakly to the Dirac-delta $\delta_{\eta_i}$ in the
semiclassical limit, for $i=1,2$.

The Monge distance defined above is invariant under the action of group
translations, namely:

\medskip

{\bf Property B.} ({\sl invariance}) Let $\alpha, \beta \in G$ and $\rho_1,
\rho_2 \in \cal{M}$. Then
\begin{equation}
D_M({R_{\beta}}^{-1} \rho_1 R_{\beta}, {R_{\beta}}^{-1} \rho_2 R_{\beta}) =
D_M(\rho_1, \rho_2) \, \text{.}
\label{invariance2}
\end{equation}
Particularly,
\begin{equation}
D_M(|\alpha \rangle, |\beta \rangle) =
D_M(|{\beta}^{-1} \alpha \rangle,|1 \rangle) \, \text{,}
\label{invariance}
\end{equation}
where $1$ is the group $G$ unit.
\medskip

The above formulae follow from the definition of the Monge distance
(\ref{mongearbmes}), and from the fact that both the measure $m$ and the metric
$d$ are translation invariant.

\subsection{Relation to other distances}

Let $\rho_1, \rho_2 \in \cal{M}$. We start from recalling the variational
formula for the trace distance (see for instance \cite{th80}).

\medskip

{\bf Proposition 3.} ({\sl variational formula for $D_{tr}$})
\begin{equation}
D_{tr}(\rho_1,\rho_2) =
\sup_{\left\| {A}\right\| \leq 1} | \operatorname*{tr} A(\rho_1-\rho_2)| \, \text{,}
\label{vartra}
\end{equation}
where the supremum is taken over all Hermitian matrices $A$ such that $\left\|
{A}\right\| \leq 1$, and the {\sl supremum norm} reads

\begin{equation}
\left\| {A}\right\| = \sup \{\left\| {Ax}\right\| : x \in {\cal {H}_N},
\left\| {x}\right\| \leq 1\} \, \text{.}
\label{norm}
\end{equation}

\medskip

Applying Proposition~1 we can prove an analogous formula for the Monge
distance

\medskip

{\bf Proposition 4.} ({\sl variational formula for $D_{M}$})
\begin{equation}
D_{M}(\rho_1,\rho_2) = \max_{L(A) \leq 1}| \operatorname*{tr} A(\rho_1-\rho_2)| \, \text{,}
\label{varmon}
\end{equation}
where the maximum is taken over all Hermitian matrices $A$ with $L(A) \leq 1$,
and

\begin{equation}
L(A)= \inf \{c : \text{there exists a $c$-Lipschitzian function} \ f:\Omega
\rightarrow {\mathbb{R}} \ \text{such that} \
A = \int_{\Omega} f(\eta) | \eta \rangle \langle \eta | dm(\eta) \} \, \text{.}
\label{Lipnor}
\end{equation}

For the proof see Appendix~B.
This proposition has a simple physical interpretation. It says that {\sl the
Monge distance between two quantum states is equal to the maximal value of the
difference between the expectation values (in these states) of observables
(Hermitian operators) some of whose P-representations are weak contractions}.
Recently, Rieffel \cite{ri99} considered the class of metrics on state spaces
which are generated by Lipschitz seminorms. If $\Omega$ is compact, then one can
show that the Monge metric $D_M$ belongs to this class.

From Propositions~3~and~4 we can also easily deduce Proposition~2.
Using Proposition~2 and the H{\"o}lder inequality for the trace (see
\cite{th80}) one can prove the following inequalities

\begin{equation}
{2 \over \Delta} D_M \leq D_{L_1} \leq N \cdot D_{HS} \leq N \cdot D_{tr}
\, \text{,}
\label{seqine}
\end{equation}
where $\Delta$ is the diameter of $\Omega$ and $N = \dim \cal{H}_N$.
On the other hand from the fact that the Monge-Kantorovich metric generates the
weak topology in the space of probability measures on $\Omega$, it follows that
$D_M(\rho_1,\rho_2) \to 0$ implies $D_{HS}(\rho_1,\rho_2) \to 0$ for every
$\rho_1, \rho_2 \in \cal{M}$. Thus the Monge metric $D_M$ and the Hilbert-Schmidt metric $D_{HS}$ generate the same topology in the space of mixed states
$\cal{M}$.

Let us emphasise here a crucial difference between our `classical' Monge
distance and the standard distances in the space of quantum states. Given any
two quantum states represented by the density matrices $\rho_1$ and $\rho_2$,
one may directly compute the trace, the Hilbert-Schmidt or the Bures distance
between them.
On the other hand, the classical distance is defined by specifying the set of
generalised coherent states in the Hilbert space. In other words, one needs to
choose a classical phase space with respect to which the Monge distance is
defined. Take for example two density operators of size $N=3$. The distance
$D_M(\rho_1,\rho_2)$ computed with respect to the $SU(2)$ coherent states and,
say, with respect to the $SU(3)$ coherent states can be different. The
simplest case of the $SU(2)$ coherent states corresponding to classical dynamics
on the sphere is discussed in the following section.

\section{Monge metric on the sphere}

\subsection{Spin coherent states representation}

Let us consider a classical area preserving map on the sphere $\Theta:S^2\to
S^2$ and a corresponding quantum map $U$ acting in an $N-$dimensional
Hilbert space ${\cal H}_N$. A link between classical and quantum mechanics
can be established via a family of spin coherent states $|\vartheta,\varphi
\rangle\in{\cal H}$ localised at points $(\vartheta,\varphi)$ of the sphere $
{\mathbb C}P^1=S^2$. The vector coherent states were introduced by Radcliffe
\cite{ra71}
and Arecchi {\sl et al.} \cite{acgt72} and their various properties are often
analysed in the literature (see e.g. \cite{zfg90,vs95}). They are related to
the $SU(2)$ algebra of the components of the angular momentum operator $
J=\{J_x, J_y, J_z\}$, and provide an example of the general group theoretic
construction of Perelomov \cite{pe86} (see Sect. III~C).

Let us choose a reference state $|\kappa\rangle$, usually taken as the maximal
eigenstate $|j,j\rangle$ of the component $J_z$ acting on ${\cal H}_N$,
$N=2j+1$, $j=1/2,1,3/2, \dots$. This state, pointing toward the
`north pole' of the sphere, enjoys the minimal uncertainty equal to $j$. Then,
the vector coherent state is defined by the Wigner rotation matrix
$R_{\vartheta,\varphi}$

\begin{equation}
|\vartheta,\varphi \rangle = \thinspace
R_{\vartheta,\varphi}|\kappa\rangle = \thinspace
(1+|\gamma|^2)^{-j} e^{\gamma J_-} |j,j\rangle \, \text{,}
\label{vecoh}
\end{equation}
where $R_{\vartheta,\varphi}
= \exp\left[{i\vartheta \left({ \cos \varphi J_x- \sin \varphi
J_y}\right)}\right]$, $J_- = J_x - iJ_y$ and $\gamma={\tan}(\vartheta/2)
e^{i\varphi}$, for $(\vartheta,\varphi) \in S^2$ (we use the spherical
coordinates).

We obtain the coherent states identity resolution in the form

\begin{equation}
\int_{S^{2}}|\vartheta ,\varphi \rangle \langle \vartheta ,\varphi |\
d\mu(\vartheta,\varphi) = {\bf I}/(2j+1) \, \text{,}
\label{resol}
\end{equation}
where the Riemannian measure $d\mu(\vartheta,\varphi) = (\sin \vartheta / 4\pi)
d\vartheta d\varphi$ does not depend on the quantum number $j$.

\smallskip

Expansion of a coherent state in the common eigenbasis of $J_z$ and $J^2$:
$|j,m\rangle$, $m=-j,\dots,+j$ (in ${\cal H}_N$)
reads

\begin{equation}
|\vartheta ,\varphi \rangle = \sum_{m=-j}^{m=j} \sin ^{j-m}
(\vartheta / 2)\cos ^{j+m}(\vartheta / 2)\exp \left({i(j-m)\varphi} \right)
\left[ \binom{2j}{j-m} \right]^{1/2}|j,m\rangle \, \text{.}
\label{thetrot}
\end{equation}
The infinite `basis' formed in the Hilbert space by the coherent states is
overcomplete. Two different $SU(2)$ coherent states overlap unless they are
directed into two antipodal points on the sphere. Expanding the coherent states
in the basis of ${\cal H}_N$ as in (\ref{thetrot}) we calculate their overlap

\begin{equation}
{|\langle \vartheta^{\prime},\varphi^{\prime}|\vartheta,\varphi \rangle|^2}
= \cos^{4j}(\Xi/2) \, \text{,}
\label{overla1}
\end{equation}
where $\Xi$ is the angle between two vectors on $S^2$ related to the coherent
states $|\vartheta, \varphi \rangle$ and $|\vartheta^{\prime},
\varphi^{\prime}\rangle$, and for $j=1/2$ it equals to the geodesic distance
(\ref{FS}).
 Thus, we have

\begin{equation}
H_{|\vartheta, \varphi \rangle \langle \vartheta, \varphi|}
(\vartheta^{\prime},\varphi^{\prime}) =
(2j+1) \cos^{4j}(\Xi/2) \, \text{.}
\label{overlaHus}
\end{equation}

This formula guarantees  that the respective
 Husimi distribution of an arbitrary
spin coherent state tends to the Dirac {\hbox{$\delta $--function}} in the
semiclassical limit $j\to\infty $.

To calculate the Monge distance between two arbitrary density matrices $\rho_1$
and $\rho_2$ of size $N$ one uses the $N=(2j+1)-$dimensional
representation of the spin coherent states $|\vartheta,\varphi\rangle$ (to
simplify the notation we did not label them by the quantum number $j$).
Next, one computes the generalised Husimi representations for both states

\begin{equation}
H_{\rho_i}(\vartheta,\varphi):=
N \cdot \langle \vartheta,\varphi|\rho_i|\vartheta,\varphi\rangle
\label{husgen1}
\end{equation}
and solves the Monge problem on the sphere for these distributions. Increasing
the parameter $j$ (quantum number) one may analyse the semiclassical properties
of the Monge distance.

It is sometime useful to use the stereographical projection of the sphere
$S^{2}$ onto the complex plane. The Husimi representation of any state $\rho$
becomes then the function of a complex parameter $z$. It is easy to see that
for any pure state $|\psi \rangle \in {\cal P}$ the corresponding Husimi
representation is given by a polynomial of $(N-1)$ order: $
W_{\psi}(z)=z^{N-1}+\sum_{i=0}^{N-2}c_{l}z^{l}=0$ with arbitrary complex
coefficients $c_{i}$. This fact provides an alternative explanation of the
equality ${\cal P}={\mathbb C}P^{N-1}$. Thus, every pure state can be uniquely
determined by the position of the $(N-1)$ zeros of $W_{\psi}$ on the complex
plane (or, equivalently, by zeros of $H_{| \psi \rangle}$ on the sphere). Such
{\sl stellar} representation of pure states is due to Majorana \cite{ma32}
and it found several applications in the investigation of quantum dynamics
\cite{ba74,pe94,lv90}. In general, the zeros of Husimi representation may be
degenerated. This is just the case for the coherent states: the Husimi function
of the state $|\vartheta ,\varphi \rangle $ is equal to zero only at the
antipodal point and the $(N-1)-$fold degeneracy occurs.
The stellar representation is used in section VI to define
a simplified Monge metric in the space of pure quantum states.

\subsection{Monge distance between some symmetrical states}

Consider two quantum states, whose Husimi distributions are invariant
with respect to the horizontal rotation. Using Proposition~4 one may found the
Monge distance between both states with the help of the Salvemini formula
({\ref{salv})

\medskip

{\bf Proposition 5.} Let $\rho_1, \rho_2 \in \cal{M}$ \ fulfil \
${H}_{\rho_i}(\vartheta ,\varphi )={\tilde{H}}_{\rho_i}(\vartheta )$ \
for $(\vartheta,\varphi) \in S^2$. Consider the normalised one-dimensional
functions $h_{\rho_i}:[0,\pi] \to \mathbb{R}^+$ given by
$h_{\rho_i}(\vartheta ):= {\tilde{H}}_{\rho_i}(\vartheta) \frac{1}{2}
\sin \vartheta$, that satisfy $\int_{0}^{\pi }h_{\rho_i}(\vartheta ) d\vartheta
=1$ for $i=1,2$.
Then

\begin{equation}
D_{M}(\rho _{1},\rho _{2}) =
\int_{0}^{\pi }\left| F_{\rho_1}(\vartheta)-F_{\rho_2}(\vartheta )\right|
d\vartheta \, \text{,}
\label{salv2}
\end{equation}
where the cumulative distributions read $F_{\rho_i}(\vartheta )=
\int_{0}^{\vartheta}h_{\rho_i}(\psi ) d\psi$.

The proof is given in Appendix~C. We use this proposition
computing the Monge distance between two arbitrary eigenstates of the operator
$J_z$ (Sect. IV F1) and the Monge distance
of some of these eigenstates from the maximally mixed state (Sect. IV F2).

\medskip

The maximally mixed state $\rho _{\ast }={\bf I}/N$ is represented by the
uniform Husimi distribution on the sphere. Thus,
${H}_{\rho _{\ast}}(\vartheta ,\varphi ) =
{\tilde {H}}_{\rho_{\ast}}(\vartheta) =1$,
$h_{\rho _{\ast }}(\vartheta )= \frac{1}{2} \sin\vartheta$, and
$F_{\rho _{\ast }}(\vartheta )= \frac{1}{2}(1-\cos \vartheta )$.

Furthermore, all the eigenstates of $J_{z}$, forming the basis $|j,m\rangle$,
possess this symmetry, and according to (\ref{thetrot}) we get
\begin{equation}
h_{|j,m\rangle \langle j,m|}(\vartheta ) = (2j+1)
\binom{2j}{j-m}
\sin^{2(j-m)+1}(\vartheta /2)\cos ^{2(j+m)+1}(\vartheta /2) \, \text{.}
\label{husjm}
\end{equation}
The formula (\ref{salv2}) enables us to compute the Monge distance between them.

\begin{figure}
 \begin{center}
\
 \vskip -0.4cm
\includegraphics[width=7.5cm,angle=0]{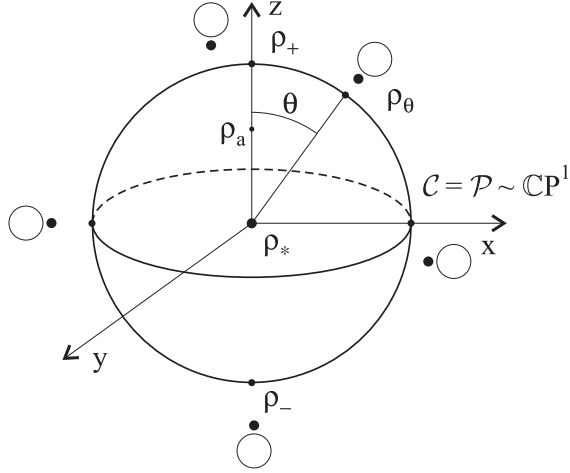}
 \end{center}
\vskip 0.0cm
\caption{Quantum states, for which we calculate Monge distances,
are denoted at the Bloch sphere corresponding to $j=1/2$.
Dots at small circles represent the position of zeros
of the Husimi function of the corresponding pure states.}
\label{f3}
\end{figure}

We introduce the notation
$\rho_+ = |j,j\rangle \langle j,j|, \
\rho_- = |j,-j\rangle \langle j,-j|$,
and
$\rho_a = a\rho_+ + (1-a)\rho_-$ for $a \in [0,1]$ (for $N=2$ these states are
represented in Fig.~3).
It follows from Proposition~1 that $H_{\rho_a} = aH_{\rho_+} + (1-a)H_{\rho_-}$,
and, consequently, $D_M(\rho_+,\rho_a) = (1-a) D_M(\rho_+,\rho_-)$, and
$D_M(\rho_a,\rho_-) = a D_M(\rho_+,\rho_-)$.

In some cases we can reduce the two-dimensional problem to the Salvemini
formula, even if it does not possess rotational symmetry.

\medskip

{\bf Proposition 6.} Let $\rho_1, \rho_2 \in \cal{M}$.
Define $F_i : [0,\pi] \times [0,2\pi] \to [0,1]$ by
$F_i(t,\varphi) = \frac{1}{2} \int_0^t H_{\rho_i}(\vartheta,\varphi)
\sin\vartheta \thinspace d \vartheta$ for $t \in [0,\pi]$, $\varphi \in
[0,2\pi]$, and $i=1,2$.
Assume that

\medskip

1. $F_1(\pi,\varphi) = F_2(\pi,\varphi)$ for all $\varphi \in [0,2\pi]$,

\medskip

2. $F_1(t,\varphi) \geq F_2(t,\varphi)$ for all $t \in [0,\pi]$ and
$\varphi \in [0,2\pi]$.

\medskip

Then
\begin{equation}
D_M(\rho_1,\rho_2) = \frac{1}{2\pi} \int_0^{2\pi} \int_0^\pi
\left( F_1(t,\varphi) - F_2(t,\varphi) \right) dt d \varphi \, \text{.}
\label{Prop6}
\end{equation}

\medskip

For the proof see Appendix~D. We use the above proposition computing both, the 
Monge distance between two arbitrary density matrices for $j=1/2$ (Sect.~IVC), 
and the Monge distance between two coherent states for arbitrary $j$ 
(Sect.~IV F3).

\subsection{Monge distances for $j=1/2$}

Let us start from the calculation of the Monge distance
between the `north pole' $\rho_+$ and a mixed state $\rho_a$ parameterised
by $a\in(0,1)$ (note that $\rho_* = \rho_{1/2}$ in this case). Computing the
distribution functions we get
$F_{\rho_a}(\vartheta)=(\sin^2\vartheta)(2a-1)/4 +(1-\cos\vartheta)/2$, while $
F_{\rho_+}(\vartheta)=F_{\rho_{a=1}}(\vartheta)$. Elementary integration
(\ref{salv2}) gives the result $D_M(\rho_+,\rho_a)=(1-a)\pi/4$. Substituting $
a=1/2$ for $\rho_*$ or $a=0$ for $\rho_-$ we get two important special
cases:
\begin{equation}
D_M(\rho_+,\rho_-)=\pi/4; \quad D_M(\rho_+,\rho_*)=D_M(\rho_-,\rho_*)=\pi/8
\, \text{.}
\label{mongj12}
\end{equation}
These three states $\rho_+$, $\rho_*$ and $\rho_-$ lay on a metric line. This
follows from the property of the distribution functions visible in Fig.~4.
They do not intersect, and therefore the area between $F_+$ and $F_-$ is
equal to the sum of two figures: one enclosed between $F_+$ and $F_*$, and
the other one enclosed between $F_*$ and $F_-$. Note, however, that the
distance $D_M(\rho_+,\rho_-)\approx 0.785$ is much smaller than the
classical distance between two poles on the sphere equal to $\pi$. Instead
of rotating the distribution $H_{\rho_+}$ by the angle $\pi$, the optimal
Monge plan consists in moving south the `sand' occupying the north pole,
along each meridian. The difference between both transformations is so large
only in this deep quantum regime, for which the distributions are very broad
and strongly overlap. As demonstrated below, this effect vanishes in the
semiclassical regime $j\to\infty$, where the semiclassical property
(\ref{clasprop}) is recovered.
\begin{figure}
 \begin{center}
\
 \vskip -2.1cm
\includegraphics[width=14cm,angle=270]{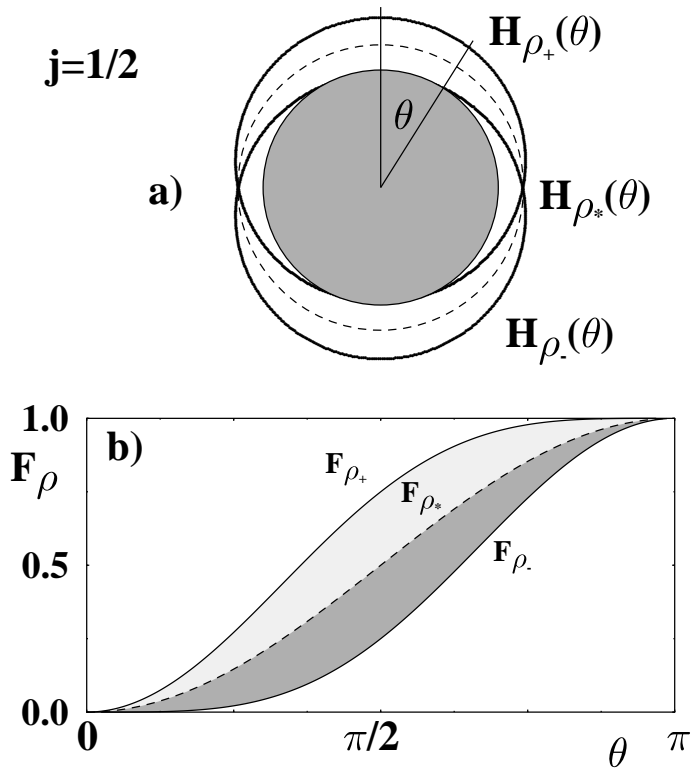}
 \end{center}
\vskip -2.5cm
\caption{a) Cross-section of the Husimi distribution $H_{\protect\rho}(
\protect\vartheta)$ for $\protect\rho_+$ (peaked at the top), for $\protect
\rho_*$ (uniform), and for $\protect\rho_-$ (peaked at the bottom) plotted
for $j=1/2$. b) Monge distance between these states may be represented as the
area between graphs of the corresponding distribution functions $F_{\protect
\rho_+}(\protect\vartheta)$, $F_{\protect\rho_*}(\protect\vartheta)$, and $
F_{\protect\rho_-}(\protect\vartheta)$. }
\label{f4}
\end{figure}

In the case $j=1/2$ all pure states are coherent, so the Monge
distance from $\rho_*$ is the same for every pure state (as illustrated in
Fig.~3). Thus, the manifold of pure states (the Bloch sphere) forms, in the
sense of the Monge metric, the sphere $S^2$ of radius $R_1=\pi/8$ centred at
$\rho_*$. All mixed states are less localised than coherent and their distance
to $\rho_*$ is smaller than $R_1$. To see this note that every mixed state can
be represented as a vector $\bf v$ in the unit ball. Using Proposition~6 and
some geometrical considerations one can find the
Monge distance between any two mixed states $\rho_1$ and $\rho_2$.
Representing them by Pauli matrices $\vec {\sigma}$
and vectors $\vec v_i$ in the Bloch ball of radius
$1/2$, namely, $\rho_i=\rho_* + {\vec \sigma} \cdot {\vec v_i}$, we
obtain\cite{SZ01} 
\begin{equation}
D_M(\rho_1,\rho_2) = \frac{\pi}{4} \, d({\vec v_1},{\vec v_2})
= \frac{\pi}{4}|{\vec v_1} - {\vec v_2}|
\, \text{,}
\label{twodim}
\end{equation}
where $d$ denotes the Euclidean metric in ${\mathbb R}^3$.
Consequently, for $j=1/2$ the Monge distance induces the
same geometry as that of the Bloch ball, as illustrated in Fig.~3.

\subsection{Monge distances for $j=1$}

In an analogous way we treat the case $N=3$. Obtained data
\begin{eqnarray}
R_1:=D_M(\rho_+,\rho_*)=D_M(\rho_-,\rho_*)=3\pi/16; \quad
D_M(\rho_+,\rho_-)=3\pi/8; \\
R_2:=D_M(\rho_*,\rho_{|0\rangle})= 1/6; \quad
D_M(\rho_+,\rho_{|0\rangle})=D_M(\rho_-,\rho_{|0\rangle})= 3\pi/16.
\label{mongj1}
\end{eqnarray}
are based on the results derived in Appendix~E (see also the next subsection)
and visualised in Fig.~5b. Note that both triples
$\{\rho_+,\rho_*,\rho_-\}$ and $\{\rho_+,\rho_{|0\rangle},\rho_-\}$ lay on
two different metric lines. Thus, in contrast to the case $j=1/2$, the two
states $\rho_+$ and $\rho_-$ are connected by several different metric lines.

\begin{figure}
 \begin{center}
\
 \vskip -0.55cm
\includegraphics[width=5cm,angle=0]{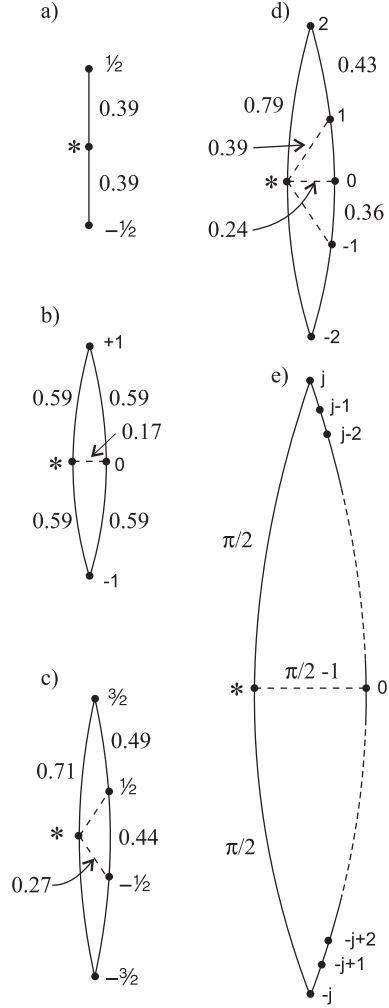}
 \end{center}
\vskip 0.2cm
\caption{Schematic map showing the eigenstates
$|j,m\rangle$ and the mixed state
$\rho_*$ for
a) $j=1/2$, b) $j=1$, c) $j=3/2$, d) $j=2$, and e) $j\to \infty$.
The symbol $*$ labelling dots represents the maximally mixed state
$\rho_*$, while numbers
$m$ denote pure states $|j,m\rangle$.
Solid lines denote the metric lines,
and the accompanying
numbers represent the approximate Monge distance between the states.
}
\label{f5}
\end{figure}

Now, consider a mixed state $\rho_m$ represented in the canonical basis by a
diagonal density matrix $\rho_m = \operatorname*{diag}(a,b,c)$, where $a+b+c=1$.
Since $\{\rho_+,\rho_{|0\rangle},\rho_-\}$ \ lay on a metric line and their
distributions are invariant with respect to the horizontal rotation, it is not
difficult to calculate the Monge distance $D_M(\rho_+,\rho_m)$ using
Proposition~5. The corresponding distribution functions do not cross, and so
$D_M(\rho_+,\rho_m) = b D_M(\rho_+,\rho_{|0\rangle}) + c D_M(\rho_+,\rho_-) =
\linebreak
\frac{3\pi}{16}(2-b-2a)$. For comparison $D_{HS}(\rho_+,\rho_m)
= \sqrt{(1-a)^2+b^2+(1-a-b)^2}$, $D_{tr}(\rho_+,\rho_m) = \sqrt{2(1-a)}$, and
$D_{Bures} = \sqrt{2(1-\sqrt{a})}$. This simple example shows that for $j=1$ the
Monge metrics induces a non-trivial geometry, considerably different from
geometries generated by any standard metric.

The Monge distance $R_{\psi}$ between any pure state $|\psi\rangle$ and the
mixed state $\rho_*$ depends only on the angle $\chi$ between two zeros of
the Husimi function located on the sphere. If the zeros are degenerated,
$\chi=0$, the state is coherent and $R_{\psi}=R_1$. The coherent states are
as much localised in the phase space, as allowed by the Heisenberg
uncertainty relation. It is therefore intuitive to expect, that out of all
pure states the coherent states are the most distant from $\rho_*$. In the
other extreme case, both zeros lay at the antipodal points, $\chi=\pi$, which
corresponds to $\rho_{|0\rangle}$, and $R_{\psi}=R_2$. In this symmetrical
case, the Husimi distribution is as delocalized as possible, and we
conjecture that for every pure state its distance to $\rho_*$ is larger than or
equal to $R_2$.

Thus, considering the Monge distance from $\rho_*$, one gets a
foliation
of the space of pure states ${\cal P} = {\mathbb C}P^2$, as shown in
Fig.~6b.
As a running parameter we may take the angle $\chi$, which describes a pure
state in the stellar representation.
This foliation is singular, since the topology of the leaves
depends to the angle.
The angle $\chi=0$ represents the sphere
$S^2$ of coherent states ($\sim SO(3)/SO(2)$), intermediate angle
 represents a generic 3D manifold 
${\mathbb R}P^3/{\mathbb Z}_2$ (a desymmetrized Stiefel manifold $\sim
SO(3)/{\mathbb Z}_2$)  of pure states of the same $\chi$, while the limiting
value $\chi=\pi$ denotes the ${\mathbb R}P^2$ ($\sim SO(3)/O(3)$) manifold
of states rotationally equivalent to $\rho_{|0\rangle}$. Similar foliations
of ${\cal P}$ discussed in other context may be  found in Bacry \cite{ba74}
and in a recent paper by  Barros e S\'{a} \cite{ba00}. For comparison in
Fig.~6a we present the foliation of ${\mathbb C}P^1$ as
regards the Monge distance from $\rho_+$.

\begin{figure}
 \begin{center}
\
 \vskip -0.4cm
\includegraphics[width=16.3cm,angle=0]{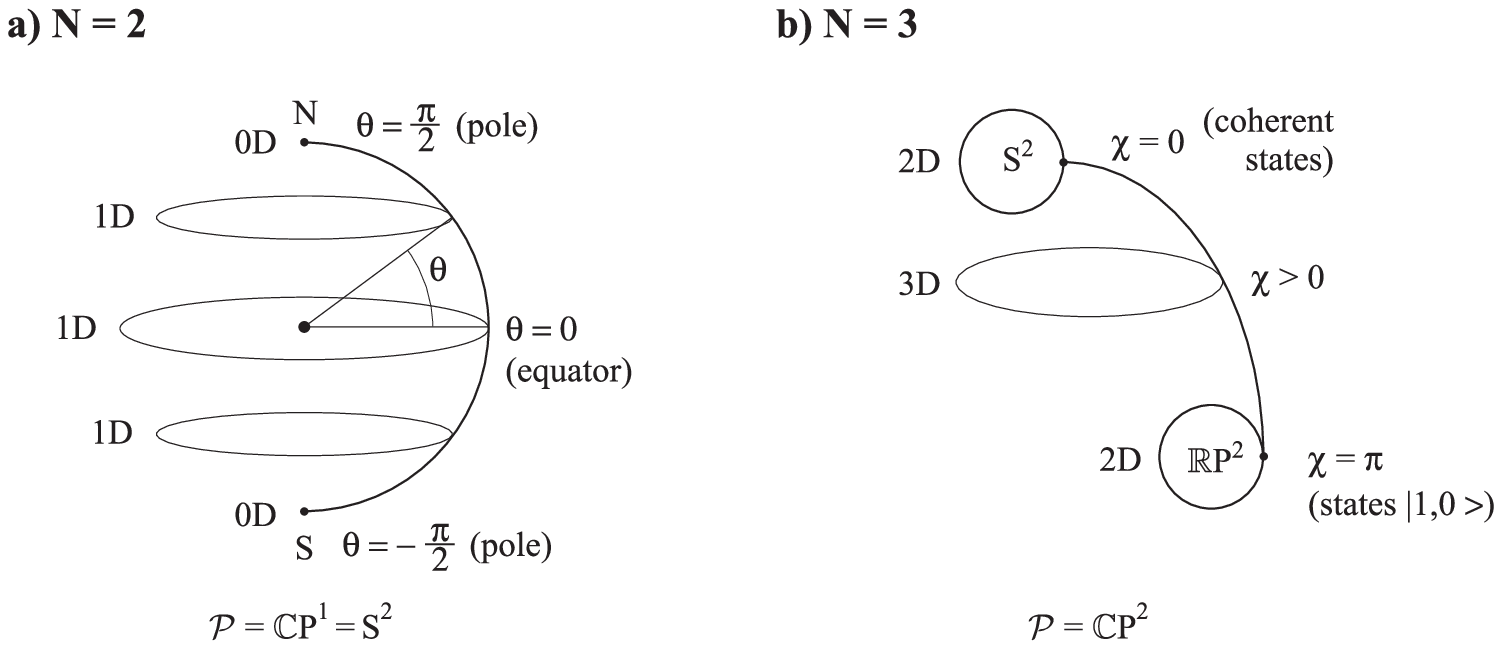}
 \end{center}
\vskip 0.05cm
\caption{Foliation of the sphere along the Greenwich
meridian  (a), foliation
of the four-dimensional space of the $N=3$ pure states along the angle $\chi$
between
both zeros of the corresponding Husimi distribution. The poles correspond to the
distinguished $2D$ submanifolds of ${\mathbb C}P^2$: the manifold of coherent
states
and the manifold of states equivalent to $|1,0\rangle$.}
\label{f6}
\end{figure}

\begin{figure}
 \begin{center}
\
 \vskip -0.8cm
\includegraphics[width=6.2cm,angle=0]{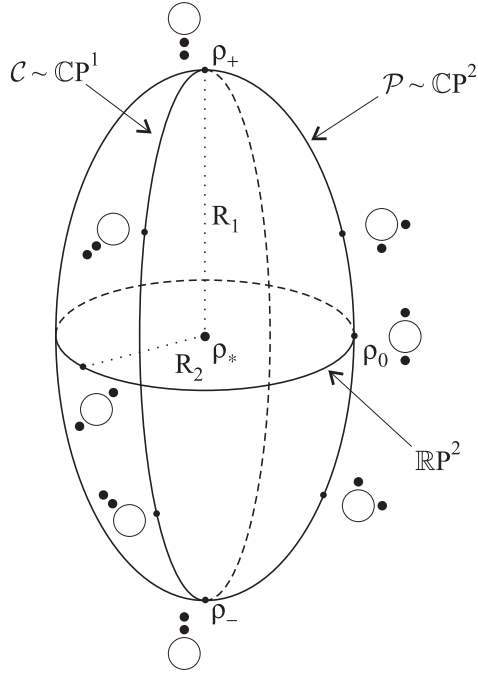}
 \end{center}
\vskip -0.2cm
\caption{Sketch of the structure of the space of the mixed states ${\cal M}$
for $j=1$ induced by the Monge metrics. Manifold of the coherent states
${\cal C}=S^2$ is represented by a circle of radius
$R_1=3\pi/16 \sim
 0.589 $ centred at $\rho_*$. Pure states isomorphic to
$\rho_{|0\rangle}$ are situated $R_2 =1/6\sim 0.166$ from $\rho_*$. Dots at
smaller circles represent the positions of zeros $z_1$ and $z_2$,
 which
determine each pure state in the stellar representation.}

\label{f7}
\end{figure}
\medskip

Since it is hardly possible to provide a plot of ${\cal M}$ revealing all
details of this non-trivial, $8$-dimensional space of mixed states, we can not
expect too much from Fig.~7, which should be treated with a pinch of salt.
As it was discussed in Sect.~II, from the point of view of the
standard metrics, the four-dimensional manifold of the pure states ${\cal P}$ is
contained in the sphere $S^7$ centered
at $\rho_*$. For the Monge metric one has $R_1>R_2$, so we
suggest to illustrate ${\cal M}$ as an $8$-dimensional full `hyper-ellipsoid'.
Pure states
$\rho_+$ and $\rho_-$ occupy its poles along the longest axis. The dashed
vertical ellipse represents the space of all coherent states ${\cal C}$, which
forms the sphere $S^2$ of radius $R_1$. Solid horizontal ellipse represents
these pure states, which are closest to $\rho_*$. This subspace may be
obtained from $|0\rangle\langle 0|$ by a three-dimensional rotation of
coordinates; topologically it is a real projective
space ${\mathbb R}P^2$. Although both ellipses do cross in the picture, both
manifolds do
not have any common points, what is easily possible in the four-dimensional
space
${\cal P}$. To simplify the identification of single pure states we added
in the picture small circles with two dark dots, which indicate their
stellar representations. In general, the states represented by points inside the
hyper-ellipsoid are mixed. However, since $\cal M$ is only a part of the hyper-
ellipsoid, not all points inside this figure do represent existing mixed states.

\begin{figure}
\begin{center}
\
 \vskip -2.3cm
\includegraphics[width=13.2cm,angle=270]{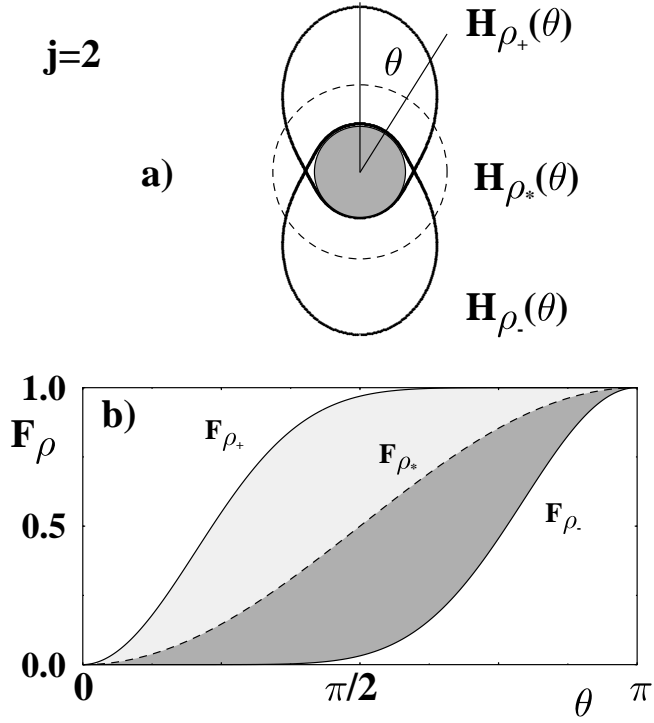}
 \end{center}
\vskip -2.5cm
\caption{As in Fig.~4 for $j=2$.}
\label{f8}
\end{figure}

\subsection{The cases: j=3/2 and j=2}

For $j=3/2$ ($N=4$) the results read in a simplified notation
$D^{(N=4)}_{3/2,1/2} = D^{(N=4)}_{-1/2,-3/2} = 5\pi/32$, $D^{(N=4)}_{1/2,-1/2} =
9\pi/64$,
$D^{(N=4)}_{3/2,*} = D^{(N=4)}_{-3/2,*} = 29\pi/128$,
and $D^{(N=4)}_{1/2,*} = D^{(N=4)}_{-1/2,*} \simeq 0.2737$.
For $j=2$ ($N=5$) one obtains
$D^{(N=5)}_{2,1} = D^{(N=5)}_{-1,-2} = 35\pi/256$,
$D^{(N=5)}_{1,0} = D^{(N=5)}_{0,-1} = 15\pi/128$,
$D^{(N=5)}_{2,*} = D^{(N=5)}_{-2,*} = 65\pi/256$,
$D^{(N=5)}_{1,*} = D^{(N=5)}_{-1,*} \simeq 0.3909$, and
$D^{(N=5)}_{0,*} = 29/120.$
Fig.~5c,d presents a schematic map of these states. Although the results are
analytical, we give their numerical approximations, which give some flavour
of the geometric structure induced by the Monge metric.

\subsection{Monge distances for an arbitrary $j$}

\subsubsection{Eigenstates of ${\bf J_z}$}

Using the formula for distribution functions $F(\vartheta)$ one may express
the distances between neighbouring eigenstates of $J_z$ for an arbitrary $j$ by
the following formula

\begin{equation}
D_M(|j,m\rangle, |j,m-1\rangle) = \pi {2 \left( N-n \right) \choose {N-n}}{2n
\choose n}2^{-2N}
\label{mongeeigen}
\end{equation}
for $m=-j+1,\dots,j$, where $N=2j+1$ and $n=j+m=1,\dots,N-1$ (for the proof see
Appendix~E1). This leads to the following asymptotic formula
\begin{equation}
D_M(|j,m\rangle, |j,m-1\rangle) \sim \frac{1}{\sqrt{n(N-n)}}
\label{mongeeigenasympt}
\end{equation}
valid for large $j$, where $n$ is defined above.
It is easy to show that that for each $j$
all the eigenstates of $J_z$ are located on one metric line. Hence we get
\begin{equation}
D_M(\rho_+,\rho_-) = D_M(|j,j\rangle, |j,-j\rangle) =
\sum_{m \thinspace = \thinspace -j+1}^{j} D_M(|j,m\rangle, |j,m-1\rangle) =
\pi \left[ 1-{\ { {2N \choose N}} }2^{1-2N} \right] \, \text{.}
\label{mongeSN}
\end{equation}

\subsubsection{Distance from $\rho_*$}

According to Property~B the distance of each coherent state from $\rho_*$ is the
same and equal to $R_1=D_M(\rho_+,\rho_*)$. This quantity may be found
explicitly for an arbitrary $N=2j+1$:
\begin{equation}
D_M(\rho_+,\rho_*)= \frac{1}{2} D_M(\rho_+,\rho_-)=
\frac{\pi}{2} \left[ 1-{\ { {2N \choose N}} }2^{1-2N} \right] \, \text{,}
\label{monjjj}
\end{equation}
which is asymptotically (for large $N$) equal to $\pi/2 - \sqrt{\pi/N}$. Such a
quantity is shown in Fig.~8 (for $j=2$) as the area between two corresponding
distribution functions. Observe that in comparison with Fig.~4 the coherent
states are more localised, and the area between steeper distribution
functions is larger. In the classical limit $N\to \infty$ we arrive at
$D_M(\rho_+,\rho_*)\to \pi/2$ and $D_M(\rho_+,\rho_-)\to \pi$. The latter result
has a simple interpretation: in this limit the coherent states become
infinitely sharp and the Monge plan consists in the rotation of the sphere by
the angle $\pi$. The three points $\rho_+$, $\rho_*$, and $\rho_-$ form another
metric line, which for $N>2$ is different from the metric line generated by
the  eigenstates of $J_Z$. Thus, for $N>2$, the metric induced by the Monge
distance is not `flat'. Moreover, for $j \in \mathbb{N}$ we have

\begin{equation}
D_M(|0\rangle\langle 0| ,\rho_*)= \sum_{k=1}^{j}
\allowbreak \frac{1}{2k+1}\frac{\left( 2k-1\right) !!}{\left(2k\right) !!} \,
\text{,}
\label{monostar}
\end{equation}
which tends to $\pi /2-1$ in the semiclassical limit $j \to \infty$ (for the
proof see Appendix~E2). It is well known that this convergence is very slow.

\subsubsection{Coherent states}

Now, let us consider two coherent states
$|\vartheta, \varphi \rangle$ and $|\vartheta^{\prime},
\varphi^{\prime}\rangle$. It follows from the rotational invariance of the Monge
metric (Property~B) that their distance depends only on $\Xi$ --
the angle between two vectors on $S^2$ representing these coherent
states, and is equal to the Monge distance between two coherent states lying on
the Greenwich meridian $\rho_+ = |0,0 \rangle \langle 0,0| $
 and $\rho_\Xi :=
|\Xi,0 \rangle \langle \Xi,0|$ (for $N=2$ the latter corresponds to the state
labelled in Fig.~3 by $\rho_{\theta}$).
We denote this distance by $C(\Xi,j) := D_M(\rho_+,\rho_\Xi)$.
Using Propositions~6 we obtain the following formula for this quantity
(for the proof see Appendix~E3):

\begin{equation}
C\left( \Xi,j\right) = \pi \, {\sin} \left(\Xi /2\right) W_{j}\left(
\sin^{2}\left(\Xi /2\right)\right)
\, \text{,}
\label{moncohstat}
\end{equation}
where $W_{j}$ is a polynomial of the form

\begin{equation}
W_{j}\left( x\right) := {{\frac{2j+1}{4^{j+1}}}}
\sum\limits_{_{\substack{0\leq u,v\\u+v<j}}}
S_{j,u,v} \, A_{u,v} \,
x^{u}\left(1-x\right)^{v} \, \text{.}
\label{moncohstat2}
\end{equation}
The symmetric coefficients $S_{j,u,v}$ are given by

\begin{equation}
S_{j,u,v} := \frac{\left( 2j \right) !}{\left( 2j-2\left(
u+v\right)-1 \right)! u! v! \left( u+v+1\right) !4^{u+v}} \, \text{,}
\label{moncohstat3}
\end{equation}
and the asymmetric coefficients $A_{u,v}$ by

\begin{equation}
A_{u,v} :=
\sum_{s=v+1}^{\infty} \frac{\binom{2s}{s}}{\left(
u+1+s\right) 4^{s}} \, \text.
\label{moncohstat3a}
\end{equation}
Note that $A_{u,v}$ can be also written as a finite sum
\begin{equation}
A_{u,v} = \frac{2^{2u+1}}{\binom{2u}{u}\left( 2u+1\right)} -
\sum_{s=0}^{v} \frac{\binom{2s}{s}}{\left(
u+1+s\right) 4^{s}} \, .
\label{moncohstat3b}
\end{equation}
The rank of $W_j$ is $\left\lfloor j-1/2\right\rfloor$, i.e., the largest
integer less than or equal to $j-1/2$. We have
$W_{1/2}\left( x\right) =\allowbreak\frac{1}{4}$
(and so $C(\Xi,1/2)=(\pi/4)\sin(\Xi/2)\sim D_{HS}(\rho_+,\rho_{\Xi})$),
$W_{1}\left( x\right) =\allowbreak\frac{3}{8}$,
$W_{3/2}\left( x\right) =\frac{1}{128}\left( 57+x\right) $,
$W_{2}\left( x\right) =\frac{5}{256}\left( 25+x\right) $, etc.

\smallskip

One can show that all the coefficients of the polynomial $W_j$ are positive.
This leads to the following simple lower and upper bounds

\begin{equation}
\pi \, C_j \, \sin \left( \Xi
/2\right)
\leq
C\left( \Xi,j\right)
\leq
\pi \, D_j \, \sin \left( \Xi
/2\right)
\, \text{,}
\label{moncohstat4}
\end{equation}
where
\begin{equation}
C_j := W_j(0) = \frac{j\left( 2j+1\right) }{2^{2j+1}}\,
 _{3}{\mathcal{F}}_{2} \left(
\left[ 3/2,1/2-j,1-j \right] ,\left[ 2,2\right] ,1\right) \, \text{,}
\label{moncohstat5a}
\end{equation}
and
\begin{equation}
D_j := W_j(1) = \frac{j\left( 2j+1\right)} {2^{2j+1}}\,
\left( 2 \cdot
 \; _{3}{\mathcal{F}}_{2} \left( \left[ 1,1/2-j,1-j\right] ,\left[ 2,3/2
\right] ,1\right) - \; _{3}{\mathcal{F}}_{2} \left( \left[ 1,1/2-j,1-j\right]
,\left[ 2,2\right] ,1\right) \right) \, \text{.}
\label{moncohstat5b}
\end{equation}
(here $_{3}{\mathcal{F}}_{2}$ stands for the generalised hypergeometric
function).
Note that $C_j \to 2/\pi$ and $D_j \to 1$ in the semiclassical limit
$j \to \infty$.
For two infinitesimally close coherent states the angle $\Xi \sim 0$ and we get
\begin{equation}
C\left(\Xi,j\right) \sim \frac{\pi}{2} \, C_j \, \Xi
\, \text{,}
\label{moncohstat6}
\end{equation}
with
\begin{equation}
C\left(\Xi,j\right) \to \Xi \ (j \to \infty) \, \text{,}
\label{moncohstat7}
\end{equation}
which agrees with Property~A.

\subsubsection{Chaotic states}

In the stellar representation the coherent states are represented by $N-1$
zeros merging together at the antipodal point on the sphere. These quantum
states are rather exceptional; a typical state has all zeros distributed all
over the sphere. It is known \cite{bbl92} that for the so-called
{\sl chaotic states} (eigenstates of Floquet operator corresponding to
classically chaotic systems) the distribution of zeros is almost uniform in
the phase space. Such states are entirely delocalized and their Husimi
distribution is close, in a sense of the $L_1$ metric, to the uniform Husimi
distribution $H_{\rho_*}$ corresponding to the maximally mixed state
$\rho_*$. One can therefore expect (applying Proposition~2) that the Monge
distance between these chaotic pure states $\rho_c$ and $\rho_*$ is small.
We conjecture that the mean value of the Monge distance $D_{M}(\rho_c,\rho_*)$
of randomly picked chaotic state $\rho_c$ from $\rho_*$ tends to $0$ in the
semiclassical limit $j \to \infty$.

\subsection {Correspondence to the Wehrl entropy and the Lieb conjecture}

In order to describe the phase space structure of any quantum state $\rho$
it is useful \cite{we91} to define the Wehrl entropy $S_{\rho}$ as
the Boltzmann--Gibbs entropy of the Husimi distribution
(\ref{husgen1})

\begin{equation}
S_{\rho} = - \int_{S^2} H_{\rho}(\theta,\varphi) \ln
[H_{\rho}(\theta,\varphi)] d\mu(\vartheta,\varphi).
\label{wehrl}
\end{equation}
It was conjectured by Lieb \cite{li78} that this quantity is minimal
for coherent states, which are as localised on the sphere as it is allowed by
the Heisenberg uncertainty relation. For partial results in the direction to
prove this conjecture see \cite{scu99,le88,sch99,gz01}. The minimum of
entropy
reads $S_{\rm min}=(N-1)/N-\ln N$, where the logarithmic term is due to
the normalisation of the Husimi distribution. It
was also conjectured \cite{le88} that the states with possibly regular
distribution of zeros on the sphere, which is easy to specify for
Pythagorean numbers $N=2,4,6,8,12,20$, are characterised by the largest
possible Wehrl entropy among all pure states.

Let us emphasise that for $N>>1$ the states
exhibiting small Wehrl entropy comparable to $S_{\rm min}$ are not
typical. In the stellar representation coherent states correspond to
the coalescence of all
$N-1$ zeros of Husimi distribution in one point. In a typical
situation all zeros are distributed uniformly on the sphere, and the
Wehrl entropy of such delocalized pure states is large. Averaging
over the natural Haar measure on the space of pure states $\cal P$
one may compute the mean Wehrl entropy $\langle S\rangle$
for the $N-$dimensional states. In slightly different context such
integration was performed in \cite{kmh88,jo91,sz98}
leading to
\begin{equation}
\left\langle S \right\rangle_{U(N)}=-\ln N+\Psi \left( N+1\right)
-\Psi \left( 2\right) \, \text{,}
\label{wehmean}
\end{equation}
where $\Psi$ denotes the digamma function, which
for natural arguments $k<n$ satisfies
$\Psi(n)-\Psi(k)=\sum_{m=k}^{n-1} 1/m$.
In the classical limit $N\to \infty$ the mean entropy tends to
$\gamma-1\sim-0.42278$ ($\gamma$ is the Euler constant),
which is close to the maximal possible Wehrl
entropy $S_{\rho_*}=0$.

The Wehrl entropy does not induce a metric in the space of quantum
states.
However, it describes the localisation of a quantum state in the
classical phase space and has some properties similar to the Monge
distance of a given state $\rho$ to the maximally
mixed state $\rho_*$ \cite{zy99b}.
In view of our results on the Monge distance, we advance
analogous conjectures, concerning the set of pure states ${\cal P}$
belonging to the $N-$dimensional Hilbert space.

\bigskip

{\bf Conjecture 1.} {\sl \ In the sense of Monge metric the coherent states
are pure states most distant from $\rho_*$. This maximal distance $R_1$ is
given by (\ref{monjjj}) and tends to $\pi/2$ for $N\to \infty$.}

\bigskip

{\bf Conjecture 2.} {\sl \ Pure states which maximise the Wehrl entropy are the
most close to $\rho_*$ in the sense of Monge metric. This minimal distance $R_2$
is equal to $1/6$ for $N=3$ and tends to $0$ for $N\to \infty$.}

\bigskip

In the analogy to the properties of the Wehrl entropy and formula
(\ref{wehmean}), one can expect that the mean distance \linebreak
$\langle R \rangle = \langle D_M(|\psi\rangle\langle\psi|,\rho_*)\rangle$
averaged over the natural measure on the manifold of pure states $\cal P$,
is close to the minimal distance $R_2$ and, for large $N$, is much smaller
than the maximal distance $R_1$. In other words, the coherent states,
distinguished by the fact of being situated in $\cal M$ as far from
$\rho_*$ as possible, are not generic. This observation is not surprising,
since ${\cal C}\sim {\mathbb C}P^1$ while ${\cal P}\sim {\mathbb C}P^{N-1}$,
but is not captured using any standard metrics in the space ${\cal M}$ of mixed
quantum states.

\bigskip

\section{Comparison of Monge and standard distances}

Results obtained for distances between several pairs of mixed states are
summarised in Table 2. Calculation of the
trace, Hilbert-Schmidt and Bures distances are performed directly from
the definitions provided in the Sect.~I.

Note that the geometry of the space $\cal M$ is well understood for $N=2$.
In the sense of the trace and the Hilbert-Schmidt metrics the set of all mixed
states has then the property of a ball contained inside the Bloch sphere: the
states $\rho_+, \rho_*$ and $\rho_-$ form a metric line.
The same statement is true for the Monge metric (see formula \ref{twodim}).
However, for the Bures metric the situation is different. As it was shown by
H{\"u}bner the set $\cal M$ has in this case the structure of a half of a $3$-
sphere \cite{hu92}, so $\rho_+, \rho_*$ and $\rho_-$ form an isosceles triangle.
However, the state $\rho_*$, located at the pole of $S^3$,
is equally distant (with respect to Bures metric) from all the pure states
${\cal P}$, which occupy the `hyper--equator' $\sim S^2$.

\medskip

\subsection{Geometry of quantum states for large $N$}

The data collected in Tab.~2 allow us to emphasise important
differences between
the geometry induced by the standard distances and the Monge distance.
From the points of view of all three of the standard metrics,
the distance $R$ between $\rho_*$ and any pure state is constant.
Therefore, in these standard geometries, the coherent
states are not distinguished in any sense in $\cal P$.

On the other hand, a `semiclassical' geometry, induced by the Monge
metric in the $(N^2-1)-$dimensional space $\cal M$,
distinguishes the space of coherent states ${\cal C}\sim S^2$.
Their Monge distance ($R_1$) from the centre $\rho_*$ is maximal. If we
try to visualise the $(2N-2)-$dimensional space of pure states ${\cal P}\sim
{\mathbb C}P^{N-1}$ as a `hyper-ellipsoid', the coherent states form the
`largest circle', represented by the dashed ellipse in Fig.~9.
There exists also a multidimensional subspace of $\cal P$,
consisting of delocalized pure states $\rho_c$,
with zeros of the corresponding Husimi function distributed
uniformly on the sphere. Such states are
situated close to $\rho_*$ with respect to the  Monge metric.
In the classical limit $N\to\infty$, their distance from $\rho_*$ ($R_2$) is
arbitrary small, so the manifold $\cal P$, almost touches
the maximally mixed state $\rho_*$. In this case, we might think
of $\cal M$ as of a full $(N^2-1)-$dimensional disk of radius $R_1\sim \pi/2$
centred at $\rho_*$, with coherent states at its edge and the pure states on
its surface. Since it is rather flat, and contains a lot of its `mass' close to
its centre, it resembles, in a sense, the Galaxy.

\hskip -0.66cm
$
\begin{tabular}{|c||c|c|c|c|}
\hline
States & $D_{tr}$ & $D_{HS}$ & $D_{Bures}$ & \multicolumn{1}{|c|}{$D_{Monge}
$} \\ \hline\hline
$(\rho _{+},\rho _{-})$ & $2$ & $\sqrt{2}$ & $\sqrt{2}$ &
\begin{tabular}{c}
$\pi (1-{\ { {2N \choose N}} }2^{1-2N}) \sim \pi - 2\sqrt{\pi/N}$ $\smallskip$
\end{tabular}
\\ \hline
$(\rho _{+},\rho _{\ast })=(\rho _{-},\rho _{\ast })$ & $2-\frac{2}{N}$ &
$\sqrt{1-\frac{1}{N}}$ & $
\sqrt{2-\frac{2}{\sqrt{N}}}$ &
$\pi (1/2-{\ { {2N \choose N}} }2^{-2N}) \sim \pi/2 - \sqrt{\pi/N}$
\\ \hline
~$\rho _{+},\rho _{\ast },\rho _{-}$~
\begin{tabular}{|c}
$N=2$ \\
$N>2$ \\
$~N\rightarrow \infty ~$
\end{tabular}
&
\begin{tabular}{c}
line \\
isosceles $\Delta $ \\
equilateral $\triangle $
\end{tabular}
&
\begin{tabular}{c}
line \\
isosceles $\Delta $ \\
isosceles $\Delta $
\end{tabular}
&
\begin{tabular}{c}
isosceles $\Delta $ \\
isosceles $\Delta $ \\
equilateral $\triangle $
\end{tabular}
& \ line \\
\hline
~$(| 0 \rangle, \rho _{\ast })$~
($j \in \mathbb{N}$)
&
$2-\frac{2}{2j+1}$ & $\sqrt{1-\frac{1}{2j+1}}$ & $
\sqrt{2-\frac{2}{\sqrt{2j+1}}}$ &
$\sum_{k=1}^{j}\allowbreak \frac{1}{2k+1}\frac{\left( 2k-1\right) !!}{\left(
2k\right) !!} \to \pi /2-1 $
\\ \hline
\begin{tabular}{c}
$|m \rangle$, $|m-1 \rangle$\\
~
\end{tabular}
&
\begin{tabular}{c}
$2$\\
~
\end{tabular}
&
\begin{tabular}{c}
$\sqrt{2}$ \\
~
\end{tabular}
&
\begin{tabular}{c}
$\sqrt{2}$ \\
~
\end{tabular}
&
\begin{tabular}{c}
$\pi {{2\left( N-n\right)} \choose {N-n}} {{2n} \choose {n}} 2^{-2N} \sim
\frac{1}{\sqrt{N-n}\sqrt{n}}$ \\
($n=j+m$) $\smallskip$
\end{tabular}
\\ \hline
$|-j\rangle \cdots |m\rangle \cdots |j\rangle $
&
$N$-dim simplex
&
$N$-dim simplex
&
$N$-dim simplex
&
line
\\ \hline
$(\rho _{+},\rho _{a})$ & $2(1-a)$ & $\sqrt{2}(1-a)$ & $\sqrt{2(1-\sqrt{a})}$
&
\begin{tabular}{c}
$2\pi (1-{\ { {2N \choose N}} }2^{1-2N})(1-a)$ $\smallskip$
\end{tabular}
\\ \hline
$(\rho _{-},\rho _{a})$ & $2a$ & $\sqrt{2}a$ &
\begin{tabular}{c}
$\sqrt{2(1-\sqrt{1-a})}$ $\smallskip$
\end{tabular}
&
\begin{tabular}{c}
$2\pi (1-{\ { {2N \choose N}} }2^{1-2N}) a$ \end{tabular}
\\ \hline
$\rho _{+},\rho _{a},\rho _{-}$ & line & line & $\triangle $ & line
\\ \hline
$(\rho _{+},\rho _{\Xi })$
\begin{tabular}{|c}
$N=2$ \\
\\
$N \geq 2$
\end{tabular}
& \begin{tabular}{c}
$2\sin (\Xi/2) $ \\
\\
$2\sqrt{1-\cos^{4j}(\Xi/2)}$
\end{tabular} &
\begin{tabular}{c}
$\sqrt{2}\sin (\Xi/2) $ \\
\\
$\sqrt{2-2\cos^{4j}(\Xi/2)}$
\end{tabular} &
\begin{tabular}{c}
$ \sqrt{2-2\cos (\Xi/2) }$ \\
\\
$\sqrt{2-2\cos^{2j}(\Xi/2)}$
\end{tabular} &
\begin{tabular}{c}
$(\pi/4)\sin(\Xi/2)$ \\
\\
$C(\Xi,j) = \pi \, {\sin} \left(\Xi /2\right) W_{j}\left( \sin^{2}\left(\Xi
/2\right)\right) $ \\
$\to \Xi$
\end{tabular}
\\ \hline
$(\rho _{-},\rho _{\Xi })$
\begin{tabular}{|c}
$N=2$ \\
\\
$N \geq 2$
\end{tabular}
& \begin{tabular}{c}
$2\cos (\Xi/2) $ \\
\\
$2\sqrt{1-\sin^{4j}(\Xi/2)}$
\end{tabular} &
\begin{tabular}{c}
$\sqrt{2}\cos (\Xi/2) $ \\
\\
$\sqrt{2-2\sin^{4j}(\Xi/2)}$
\end{tabular} &
\begin{tabular}{c}
$ \sqrt{2-2\sin (\Xi/2) }$ \\
\\
$\sqrt{2-2\sin^{2j}(\Xi/2)}$
\end{tabular} &
\begin{tabular}{c}
$(\pi/4)\cos(\Xi/2)$ \\
\\
$C(\Xi,j) = \pi \, {\cos} \left(\Xi /2\right) W_{j}\left( \cos^{2}\left(\Xi
/2\right)\right)$ 
\\
$\to \pi - \Xi$
\end{tabular}
\\ \hline
\end{tabular}
$

\medskip

Table 2. Standard distances (trace, Hilbert-Schmidt and Bures)
versus the Monge distance for various quantum states in $N=2j+1$
dimensions - pure states: the coherent states
$\rho_{\Xi}=|\Xi\rangle \langle \Xi|$, $\rho_{+}$, $\rho_{-}$, and the
eigenstates $|m\rangle$ of $J_z$, and mixed states: $\rho_a$, defined in
Sect.~IVB, and the maximally mixed state $\rho_*$. For the Monge distance we
give semiclassical asymptotics ($N \to \infty$). The polynomials $W_j(x)$ are
given by formula (\ref{moncohstat2}). \bigskip

\subsection{Dynamical properties }

As mentioned in Sect.~II the standard distances are preserved by
the unitary dynamics (see formula (\ref{invar})). Analogous relation is true for
the Monge distance only for some special cases,
e.g., for simple rotations $U=\exp(ia J_z)$ which preserve the coherence.
In general, however, the Monge distance
is not conserved
\begin{equation}
D_M(\rho _{1},\rho _{2}) \ne D_M(\rho _{1}^{\prime },\rho
_{2}^{\prime }) \, \text{.}
\label{inva2}
\end{equation}
Vaguely speaking, during the rotation of the `hyper-ellipsoid',
depicted in Fig.~9, a kind of contraction occurs, so
the Monge distance changes during the unitary time evolution (and hence is not a monotone metric). Since in the classical limit
the distance between coherent states tends to the classical distance
on the sphere, we suggested \cite{zsw93} to study
the time evolution of the Monge distance $D_M(t)$ between two
neighbouring coherent states.
The quantity $\lambda(t)=\lim _{D_M(0)\to 0}(\ln[D_M(t)/D_M(0)])/t$
characterises indeed
the stability of the quantum system. To get a closer analogy with the
classical Lyapunov exponent one should then perform the limit
$t\to\infty$. However, for longer times, both vector coherent states
become delocalized (under the assumption of a generic evolution operator
$U$), and their distance to $\rho_*$ becomes small. Therefore,
after some time $t_r$, the distance $D_M(t)$ starts to decrease, so
instead of analysing $\lim_{t\to \infty}\lambda(t)$ (which always tends
to zero), one needs to relay on a finite times quantity $\lambda(t_r)$
\cite{zsw93,hwz92}.

\begin{figure}
\begin{center}
\
 \vskip -0.7cm
\includegraphics[width=5.2cm,angle=0]{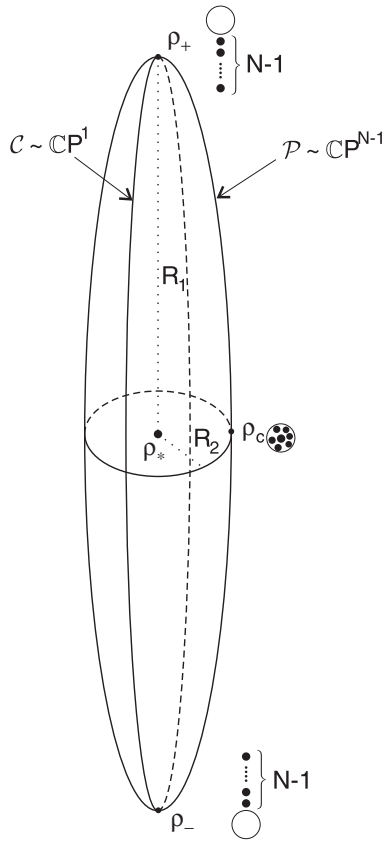}
 \end{center}
\vskip -0.1cm
\caption{Sketch of the space
${\cal M}$ in the semiclassical regime $j >> 1$.
In the limit $j \to \infty$ the larger radius $R_1$
tends to $\protect\pi/2$, while the smaller $R_2\to 0$.
Dots and small circles show the corresponding pure states
in the stellar representation.
}
\label{f9}
\end{figure}

\medskip
\subsection{Delocalisation and decoherence}
As mentioned above, the localisation of a given pure state $|\phi\rangle$
in the classical phase space is reflected by its large Monge distance from
$\rho_*$. In an analogous way one may characterise the properties of a given
Hamiltonian $H$ or a unitary Floquet operator $F$ by the mean distance of its
eigenstates $|v_i \rangle, \ i=1,\dots,N$, from the maximally mixed
state. Such a quantity, $\gamma := \sum_{i=1}^N D_M(|v_i\rangle
\langle v_i|,\rho_*)]/N$, indicates the average localisation of the
eigenstates, relevant to distinguish between integrable and chaotic quantum
dynamics \cite{ha91}. It might be thus interesting to find unitary operators
$F_{\min}$ and $F_{\max}$, for which the mean distance
$\gamma$ achieves the smallest (the largest) value.

Physical systems coupled to the environment suffer decoherence.
The density matrix of a given system tends to be diagonal in the eigenbasis of
the Hamiltonian $H_I$, which describes the interaction with the environment
\cite{zu81}. In the simplest case, $N=2$, the decoherence may be visualised
as an orthogonal projection into an axis determined by $H_I$.
For example, if $H_I$ is proportional to $J_z$, it is just the $z$ axis,
which joints both poles of the Bloch sphere.

In the general case of arbitrary $N$, there exists an $(N-1)-$dimensional
simplex $\cal I$ of density matrices diagonal in the eigenbasis of $H_I$.
Decoherence consists thus in projecting of the initial state into $\cal
I$. In a generic case of a typical interaction the eigenstates of $H_I$
are delocalized and their Monge distance from $\rho_*$ is small. On the other
hand, the typical coherent states are located far away from $\cal I$, in the
sense of the Monge metric. One can therefore expect, that the Monge distance of
a given quantum state from $\cal I$ contains the information concerning the
speed of decoherence. It is known that among all pure states the decoherence of
the coherent states is the slowest \cite{zu98}.

Moreover, the speed of decoherence of a Schr{\"o}dinger cat-like pure
state, localised at two different classical points $x$ and $x'$,
in a generic case depends on their distance in the classical phase space.
Consider now a coherent superposition
$|\psi\rangle = (|\alpha\rangle +|\beta\rangle)/ \sqrt{2}$ of
arbitrary two quantum pure states. The Monge distance between them,
$D_M(|\alpha\rangle,|\beta\rangle)$ might be thus used to characterise
the speed of the decoherence of the cat--like state $|\psi\rangle$.

\section{Simplified Monge distance between pure states}

\subsection{Definition}

With help of the stellar representation \cite{ma32,lv90,pe94} we may link
any pure state $|\varphi\rangle$ of the $N$-dimensional Hilbert
space to a singular distribution $f_{\varphi}(x)$ containing $(N-1)$ delta
peaks placed in the zeros $x_i$ of the corresponding Husimi function
$H_{|\varphi\rangle\langle \varphi |} (x)$, where $x \in S^2$,
\begin{equation}
|\varphi\rangle \to f_{\varphi}(x):={1 \over N-1} \sum_{i=1}^{N-1}
\delta(x-x_i). \label{stelf}
\end{equation}
The zeros $x_i$ may be degenerated. For any
 coherent state all $(N-1)$ zeros cluster at the antipodal point, so
$|\alpha\rangle$ is represented by $f_{\alpha}(x)=\delta(x-{\bar{\alpha}})$.

The {\sl simplified Monge distance} between any pure states $|\varphi\rangle$
and $|\psi\rangle$ is defined as the Monge distance (\ref{mongearbmes})
between the corresponding distributions (\ref{stelf})
\begin{equation}
D_{sM}(|\varphi\rangle, |\psi\rangle) := D_M(f_{\varphi},f_{\psi}).
\label{simpmong}
\end{equation}
It may be also called {\sl discrete Monge distance}, since it corresponds to
a discrete Monge problem, which may be effectively evaluated numerically
by means of the algorithms of linear programming \cite{wc81}. This contrasts
the original definition (\ref{Monge-def}), for which one needs
to solve the two dimensional Monge problem for {\sl continuous} Husimi
distributions.

Clearly, in the space of pure quantum states
both Monge distances are related. This fact becomes more transparent, if
one realises that (\ref{simpmong}) is equal to the Monge distance
between the related distributions
${\bar f}_{\varphi}:= [\sum_{i=1}^{N-1} \delta_{{\bar x}_i}]/(N-1)$ and 
${\bar f}_{\psi}:= [\sum_{i=1}^{N-1} \delta_{{\bar y}_i}]/(N-1)$, 
where $y_1,\dots,y_{N-1}$ are zeros of the Husimi distribution 
$H_{|\psi\rangle\langle\psi|}$, and points $x_i$ and ${\bar x}_i$ (resp. $y_i$ 
and ${\bar y}_i$) are antipodal on the sphere (note that the bar does not denote 
here the complex conjugation).
The distributions ${\bar f}_{\varphi}$ and ${\bar f}_{\psi}$ may be considered 
as a discrete, $(N-1)$-points approximation of the continuous Husimi 
distributions $H_{|\varphi\rangle\langle\varphi|}$ and 
$H_{|\psi\rangle\langle\psi|}$.

Since any coherent state is represented by a single Dirac delta,
${\bar f}_{\alpha}(x)= \delta (x-\alpha)$, the semiclassical condition
(\ref{clasprop}) is exact for any dimension $N$
\begin{equation}
 D_{sM}(|\eta_1\rangle ,|\eta_2\
\rangle ) = d(\eta_1,\eta_2).
\label{clasprop3}
\end{equation}
Thus for $N=2$ the discrete Monge distance, $D_{sM}$, is equal to the
Fubini-Study distance (\ref{FS}), (in this case, the Riemannian distance $d$ on
the sphere), while the continuous Monge metric, $D_M$, is proportional to the
Hilbert-Schmidt distance (\ref{HS}), (in this case, the Euclidean distance
along the cord inside the sphere). At small distances both geometries
coincide (the `flat earth' approximation).

\subsection{Eigenstates of ${\bf J_z}$}

In stellar representation the state $|j,m\rangle$
is described by $j+m$ zeros at the south pole and $j-m$ zeros
at the north pole. Thus the distribution $f_{j,m}(x)$ consists of two
delta peaks, $\pi$ apart, and it is straightforward to obtain the following
general result 
\begin{equation}
 D_{sM}(|j, m\rangle ,|j, m'\rangle ) ={\pi \over 2j} |m-m'|.
\label{simp1}
\end{equation}
In particular $D_{sM}(|j, j\rangle ,|j, -j \rangle )=\pi
=d(\alpha,{\bar \alpha})$.
 The zeros of the Husimi function of the eigenstates of the
operators $J_y$ and $J_x$ are located at the equator at the distance $\pi /2$
from both poles. Thus
\begin{equation}
 D_{sM}(|j, m\rangle_z ,|j, m'\rangle_y)=
D_{sM}(|j, m\rangle_z ,|j, m''\rangle_x)
 ={\pi \over 2}
\label{simp2}
\end{equation}
for any choice of quantum numbers $m, m'$ and $m''$.

\subsection{Random chaotic states}

Eigenstates of classically chaotic dynamical systems may be
described by random pure states \cite{ha91}.
Expansion coefficients of a chaotic state $|\psi_c\rangle$
in an arbitrary basis may be
given by a vector of a random unitary matrix,
distributed according to the Haar measure on $U(N)$.
Zeros of the corresponding Husimi
representation are distributed uniformly on the entire sphere \cite{lv90},
(with the correlations between them given by Hannay \cite{ha96}).
This fact allows one to compute the average distance of a random state to any
coherent state
\begin{equation}
 D_{sM}(|\alpha \rangle ,|\psi_c\rangle )=
{1\over 2} \int_0^{\pi} \Xi \sin \Xi \ d\Xi =
 {\pi \over 2}.
\label{simp3}
\end{equation}
In a similar way we get the average distance to the eigenstates of $J_z$
\begin{equation}
 D_{sM}(| j,m \rangle ,|\psi_c\rangle )=
 \chi \sin \chi + \cos\chi,
\quad
{\rm where}
\quad
\chi ={m\pi \over 2j}.
\label{simp4}
\end{equation}
It admits the smallest value equal to unity for $m=\chi=0$, while the
largest value is obtained for $m=\pm j$, for which the above formula
reduces to (\ref{simp3}).

Let us divide the sphere into $N$ cells of diameter proportional to
$\sqrt{N}$. Consider two different uncorrelated random states
$|\psi_{c}\rangle$ and $|\phi_{c}\rangle$.
Uniform distribution of zeros implies that there will be on average one zero
in each cell and the distance between the corresponding zeros of both states
is of order of $\sqrt{N}$. Thus their simplified Monge distance
vanish in the semiclassical limit,
\begin{equation}
 D_{sM}(|\psi_c \rangle ,|\phi_c\rangle )\approx {1 \over N} {N \over
\sqrt{N}} \sim N^{-1/2} \to 0 \quad (N \to \infty).
\label{simp5}
\end{equation}
Thus in the space of pure quantum states the simplified, discrete Monge
metric $D_{sM}$ displays several features of the original, continuous Monge
metric $D_M$.

\section{Concluding Remarks}

In this paper we analysed the properties of the set of all mixed states
constructed of the pure states belonging to the $N-$dimensional Hilbert space.
The structure of this set is highly non-trivial due to the existence of the
density matrices with degenerated spectra. Each spectrum may be represented by a
point in the $(N-1)-$dimensional simplex.
In a generic case of a nondegenerate spectrum (point located in the interior of
the simplex) this set has a structure of $[U(N)/(U(1))^N]\times G_N$.
However, there exist all together $2^{N-1}$ parts of the asymmetric
simplex of eigenvalues, all but one corresponding to its boundaries. These
boundary points, representing various kinds of the degeneration of the spectrum,
lead to a different local structure of the set of mixed states.

Standard metrics in the space of quantum states are not related with the metric
structure of the corresponding classical phase space. To establish such a link
we used vector coherent states, localised in a given region of the
sphere, which plays a role of the classical phase space.
Each quantum state may be then uniquely represented by its Husimi
distribution, which carries the information concerning its localisation
in the classical phase space. We proposed to measure
the distance between two arbitrary quantum states by the Monge distance between
the corresponding Husimi distributions. Therefore, to compute this distance, one
has to solve the Monge problem on the sphere. Thus, unexpectedly, a motivation
stemming from quantum mechanics leads us close to the original Monge problem of
transporting soil on the Earth surface. Even if the exact solution of the Monge 
problem is not accessible we can use either lower and upper bounds for the Monge 
distance (definition (\ref{monge}), Propositions~2 and 4), or numerical 
algorithms based on the idea of approximation of continuous distributions by 
discrete ones. These techniques lead to general methods of computing the Monge 
distance on the sphere (Propositions~5 and 6), as well as to concrete results we 
obtained in this paper (Sect.~IVC,D,E,F and Sect.~VA).  

The Monge distance induces a non-trivial geometry in the space
of mixed quantum states. For $N=2$ it is consistent with the geometry of the
Bloch ball induced by the Hilbert-Schmidt or the trace distance. For larger $N$
it distinguishes the coherent states, which are as localised in the phase
space, as it is allowed by the Heisenberg uncertainty principle.
These states, laying far away from the most mixed state $\rho_*$,
are not typical. The vast majority of pure quantum states are localised
in vicinity of $\rho_*$ in the sense of the Monge metric.
The Hilbert-Schmidt distance between a given state $\rho$ and $\rho_*$ may be
used to measure its degree of mixing. On the other hand, the Monge distance
$D_{M}(\rho,\rho_*)$ provides information concerning
the localisation of the state $\rho$ in the classical phase space.

A similar geometry in the space of pure quantum states is
induced by the simplified Monge metric $d_{sM}$.
It is defined by the Monge distance between the $(N-1)$--points
discrete approximations to the Husimi representation generated by the
stellar representation of pure states. This version of the Monge distance
may be easily evaluated numerically by means of the
algorithms of linear programming \cite{wc81}. Therefore it might be used
to study the divergence of initially closed pure states subjected to
unitary dynamics and to define a quantum analogue
of the classical Lyapunov exponent \cite{zsw93,wzs98}.
 Moreover, this metric may be useful
in an attempt to prove the Lieb conjecture:
it suffices to show that for any pure state $|\psi\rangle$
the Wehrl entropy decreases along the line
joining $|\psi\rangle$ with the closest coherent state.

In contrast with the standard distances, the both Monge distances are not
invariant under an arbitrary unitary transformation. This resembles the
classical situation, where two points in the phase space may
drift away under the action of a given Hamiltonian system.
In a sense, the Monge distance in the space of quantum states
enjoys some classical properties. Several classical quantities emerge in the
description of quantum systems. We believe, accordingly, that
the concept of the Monge distance between quantum states
might be useful to elucidate various aspects of the
quantum--classical correspondence.

\section{Acknowledgements}

K.~{\.Z}. would like to thank I.~Bengtsson for fruitful discussions and a
hospitality in Stockholm. It is a pleasure to
thank H.~Wiedemann for a collaboration at the early stage of this project and to
S.~Cynk, P.~Garbaczewski, Z.~Pogoda, and P.~Slater for helpful comments.
Travel grant by the European Science Foundation under the
programme {\sl Quantum Information} (K.{\.Z}) and financial support by Polish
KBN grant no 2 P03B 07219 are gratefully acknowledged.

\appendix

\section{Proof of Proposition~2}

Applying Proposition~1 we see that it suffices to prove the inequality

\begin{equation}
\left|\int_{\Omega} f(x)(Q_1(x)- Q_2(x))dm(x) \thinspace \right| \leq (\Delta/2)
D_{L_1}(Q_1,Q_2)
\label{appendixA1}
\end{equation}
for every weak contraction $f:\Omega \to \mathbb{R}$.
For such $f$ we see at once that $(\max f - \min f) \leq \Delta$. Let us
consider a function $g:\Omega \to \mathbb{R}$ defined by the formula $g(x) =
f(x) - \min f - \Delta/2$ for $x \in \Omega$. Clearly $|g| \leq \Delta/2$.
Finally, we get

\begin{equation}
\left| \int_{\Omega} f(x)(Q_1(x)- Q_2(x))dm(x) \thinspace \right| =
\left| \int_{\Omega} g(x)(Q_1(x)- Q_2(x))dm(x) \thinspace \right|
\leq (\Delta/2) D_{L_1}(Q_1,Q_2),
\label{appendixA2}
\end{equation}
which completes the proof.

\section{Proof of Proposition 4}

Let $\rho_1, \rho_2 \in \cal{M}$. We denote the set of all contractions
($c-$Lipschitzian functions with $c \leq 1$) $f: \Omega \to \mathbb{R}$ by
$Lip_1$. We have

\begin{eqnarray}
D_{M}(\rho_1,\rho_2) = D_M \left({H_{\rho_1},H_{\rho_2}} \right) =
\nonumber \\
\max_{f \in Lip_1} \left|\int_{\Omega} f(\eta)( H_{\rho_1}(\eta)-
H_{\rho_2}(\eta))dm(\eta) \thinspace \right| =
\nonumber \\
\max_{f \in Lip_1} \left| \int_{\Omega} f(\eta) \langle \eta | (\rho_1 - \rho_2)
| \eta \rangle dm(\eta) \thinspace \right| =
\nonumber \\
\max_{f \in Lip_1} \left| \operatorname*{tr} \left({ \int_{\Omega} f(\eta) |\eta \rangle \langle
\eta | dm(\eta) \thinspace (\rho_1 - \rho_2) }\right) \right| =
\nonumber \\
\max \left\{ | \operatorname*{tr} A(\rho_1-\rho_2)| : A - \text{Hermitian and} \
A = \int_{\Omega} f(\eta) | \eta \rangle \langle \eta | dm(\eta)
\ \text{for some} \ f \in Lip_1 \right\} =
\nonumber \\
\max_{L(A) \leq 1}| \operatorname*{tr} A(\rho_1-\rho_2)|
\, \text{.}
\label{appendixB1}
\end{eqnarray}

\section{Proof of Proposition~5}

We start from two simple lemmas on weak contractions on the sphere. In the
sequel, $d$ denotes the Riemannian metric on $S^2$.

\medskip

{\bf Lemma 1.}
Let $f:[0,\pi] \to \mathbb{R}$ be a weak contraction.
Define $\widetilde{f}: S^2 \to \mathbb{R}$ by the formula
\begin{equation}
\widetilde{f}(\vartheta,\varphi) = f(\vartheta)
\label{lemma1}
\end{equation}
for $(\vartheta,\varphi) \in S^2$.
Then $\widetilde{f}$ is a weak contraction.
\smallskip

{\bf Proof of Lemma 1.}
Let $(\vartheta_1,\varphi_1), (\vartheta_2,\varphi_2) \in S^2$. Applying
spherical triangle inequality we get \linebreak
$|\widetilde{f}(\vartheta_1,\varphi_1)-\widetilde{f}(\vartheta_2,\varphi_2) | =
|f(\vartheta_1) - f(\vartheta_2)| \leq |\vartheta_1 - \vartheta_2| \leq
d((\vartheta_1,\varphi_1), (\vartheta_2,\varphi_2))$.
\medskip

{\bf Lemma 2.}
Let $G:S^2 \to \mathbb{R}$ be a weak contraction.
Define $\widetilde{G}:[0,\pi] \to \mathbb{R}$ by the formula
\begin{equation}
\widetilde{G}(\vartheta) = \frac{1}{2\pi} \int_0^{2\pi} G(\vartheta,\varphi)
d \varphi
\label{lemma2}
\end{equation}
for $\vartheta \in [0,\pi]$.
Then $\widetilde{G}$ is a weak contraction.
\smallskip

{\bf Proof of Lemma 2.}
Let $\vartheta_1, \vartheta_2 \in [0,\pi]$. We have
$d((\vartheta_1,\varphi),(\vartheta_2,\varphi)) = |\vartheta_1 - \vartheta_2|$.
Hence
$|\widetilde{G}(\vartheta_1) - \widetilde{G}(\vartheta_2)| =
(1/2\pi) |\int_0^{2\pi} (G(\vartheta_1,\varphi) - G(\vartheta_2,\varphi))
d\varphi|
\leq
(1/2\pi) \int_0^{2\pi} d((\vartheta_1,\varphi),(\vartheta_2,\varphi)) d\varphi
\leq |\vartheta_1 - \vartheta_2|$.

\bigskip

{\bf Proof of formula (\ref{salv2}).}

It follows from Proposition~1 that

\begin{eqnarray}
D_{M}(\rho_1,\rho_2) = D_M \left({H_{\rho_1},H_{\rho_2}} \right) =
\nonumber \\
\max_{f \in Lip_1(S^2)} \left|\int_0^{\pi} \int_0^{2\pi} f((\vartheta,\varphi))
(H_{\rho_1}(\vartheta,\varphi)-
H_{\rho_2}(\vartheta,\varphi))dm((\vartheta,\varphi)) \thinspace \right| =
\nonumber \\
\max_{f \in Lip_1(S^2)} \left|\int_0^{\pi} \left( \int_0^{2\pi}
f((\vartheta,\varphi))(\tilde{H}_{\rho_1}(\vartheta)-
\tilde{H}_{\rho_2}(\vartheta))\sin{\vartheta}/{4\pi} \thinspace d\vartheta
\right) d\varphi \thinspace \right|
\label{appendixC3}
\end{eqnarray}

From the Salvemini formula ({\ref{salv}), Proposition~1, the above Lemma~1, and
formula (\ref{appendixC3}) we deduce that

\begin{eqnarray}
\int_{0}^{\pi }\left| F_{\rho_1}(\vartheta)-F_{\rho_2}(\vartheta )\right|
d\vartheta =
\nonumber \\
\max_{f \in Lip_1([0,\pi])} \left|\int_0^{\pi}
f(\vartheta)\left((h_{\rho_1}(\vartheta) - h_{\rho_2}(\vartheta) \right)
d\vartheta
\thinspace \right| =
\nonumber \\
\max_{f \in Lip_1([0,\pi])} \left|\int_0^{\pi} \left(
\frac{1}{4\pi}\int_0^{2\pi}
\tilde{f}((\vartheta,\varphi))(\tilde{H}_{\rho_1}(\vartheta)-
\tilde{H}_{\rho_2}(\vartheta))\sin{\vartheta} \thinspace d\vartheta
\right) d\varphi \thinspace \right| \leq
\nonumber \\
\max_{f \in Lip_1(S^2)} \left|\int_0^{\pi} \left(
\frac{1}{4\pi}\int_0^{2\pi}
f((\vartheta,\varphi))(\tilde{H}_{\rho_1}(\vartheta)-
\tilde{H}_{\rho_2}(\vartheta))\sin{\vartheta} \thinspace d\vartheta
\right) d\varphi \thinspace \right| =
\nonumber \\
D_{M}(\rho_1,\rho_2) \,
\text{.}
\label{appendixC4}
\end{eqnarray}

On the other hand, applying formula (\ref{appendixC3}), the above Lemma~2,
Proposition~1, and the Salvemini formula ({\ref{salv}) we get

\begin{eqnarray}
D_{M}(\rho_1,\rho_2)=
\nonumber \\
\max_{G \in Lip_1(S^2)} \left|\int_0^{\pi} \left( \int_0^{2\pi}
G((\vartheta,\varphi))(\tilde{H}_{\rho_1}(\vartheta)-
\tilde{H}_{\rho_2}(\vartheta))\sin{\vartheta}/{4\pi} \thinspace d\vartheta
\right) d\varphi \thinspace \right| =
\nonumber \\
\max_{G \in Lip_1(S^2)} \left|\int_0^{\pi}
\tilde{G}(\vartheta)\left((H_{1}(\vartheta)-H_{2}(\vartheta) \right)
\sin{\vartheta} \thinspace d\vartheta \thinspace \right| \leq
\nonumber \\
\max_{g \in Lip_1([0,\pi])} \left|\int_0^{\pi}
g(\vartheta)\left(H_{1}(\vartheta)
\sin{\vartheta} - H_{2}(\vartheta) \sin{\vartheta} \right) d\vartheta \thinspace
\right| =
\nonumber \\
\int_{0}^{\pi }\left| F_{1}(\vartheta)-F_{2}(\vartheta )\right| d\vartheta \,
\text{,}
\label{appendixC5}
\end{eqnarray}
which establishes the formula.

\section{Proof of Proposition~6}

Put

\begin{equation}
\widetilde{D}_M(\rho_1,\rho_2) := \frac{1}{2\pi} \int_0^{2\pi} \int_0^\pi
\left( F_1(t,\vartheta) - F_2(t,\vartheta) \right) \ dt \ d \vartheta \, .
\label{appendixD1}
\end{equation}

Let $\varphi\in\left[ 0,2\pi\right] $ and $i=1,2$. Integrating by parts we
get$\label{q}$

\begin{equation}
\int\left( -t\right) \frac{1}{2}H_{\rho_{i}}\left( t,\varphi\right) \sin
t \ dt=-tF_{i}\left( t,\varphi\right) +\int F_{i}\left( t,\varphi\right)
dt \,
\text{.}
\label{AppendixD2}
\end{equation}
Hence

\begin{equation}
\int\nolimits_{0}^{\pi}\left( F_{1}\left( t,\varphi\right) -F_{2}\left(
t,\varphi\right) \right) dt=\int\nolimits_{0}^{\pi}\left( -t\right)
\frac{1}{2}\left( H_{\rho_{1}}\left( t,\varphi\right) -H_{\rho_{2}}\left(
t,\varphi\right) \right) \sin t \ dt+\left. t\left( F_{1}\left(
t,\varphi\right) -F_{2}\left( t,\varphi\right) \right) \right|
_{t=0}^{t=\pi} \,
\text{.}
\label{AppendixD3}
\end{equation}
According to assumption (1) the last term is equal to $0$. Thus, applying
Proposition 1 to $f:S^{2}\rightarrow\mathbb{R}$ given by $f\left(
t,\varphi\right) =-t$, for $t\in\left[0,\pi\right]$, $\varphi\in\left[
0,2\pi\right]$, we get

\begin{equation}
\widetilde{D}_{M}\left( \rho_{1},\rho_{2}\right) =\int\nolimits_{0}^{2\pi
}\int\nolimits_{0}^{\pi}\left( -t\right) \frac{1}{2}\left( H_{\rho_{1}
}\left( t,\varphi\right) -H_{\rho_{2}}\left( t,\varphi\right) \right)
\sin t \ dt \ d\varphi\leq D_{M}\left( \rho_{1},\rho_{2}\right) \,
\text{.}
\label{AppendixD4}
\end{equation}

On the other hand consider the following transformation of the density
$H_{\rho_{1}}$ into $H_{\rho_{2}}$: we transport the `mass' along each
meridian separately (it is feasible due to assumption (1)) and then we join all
the transformations together. Applying the Salvemini formula (\ref{salv}) to
each meridian ($\varphi\in\left[ 0,2\pi\right] $), averaging the results
over $\left[ 0,2\pi\right] $, and finally using assumption (2) we get

\begin{equation}
D_{M}\left( \rho_{1},\rho_{2}\right) \leq\frac{1}{2\pi}\int\nolimits_{0}
^{2\pi}\int\nolimits_{0}^{\pi}\left| F_{1}\left( t,\varphi\right)
-F_{2}\left( t,\varphi\right) \right| \ dt \ d\varphi=\widetilde{D}_{M}\left(
\rho_{1},\rho_{2}\right) \,
\text{,}
\label{AppendixD5}
\end{equation}
which completes the proof.

\section{Derivation of the Monge distances for some interesting cases}

\subsection{Derivation of formula (\ref{mongeeigen})}

Let $j\in\mathbb{N}$ and $m=-j+1,\dots,j$. We put $N=2j+1$ and $n=j+m$.
Applying Proposition 5 and then the substitution $u=\cos^{2}\left(
\vartheta/2\right)$ we get

\begin{eqnarray}
D_{M}(|j,m\rangle,|j,m-1\rangle) & =\int_{0}^{\pi}\left| \int_{0}
^{y} G\left( n,N,\cos^{2}\left( \vartheta/2\right) \right)
-G\left( n-1,N,\cos^{2}\left( \vartheta/2\right) \right) \left(
\sin\allowbreak\allowbreak\allowbreak\vartheta\right) /2d\vartheta\right|
dy
\nonumber \\
& =\int_{0}^{\pi}\left| \int_{\cos^{2}\frac{y}{2}}^{1}G\left( n,N,u\right)
-G\left( n-1,N,u\right) du\right| dy \, \text{,}
\label{appendixE1}
\end{eqnarray}
where $G\left( n,N,u\right) :=N\binom{{N-1}}{n}u^{n}\left( \,1-u\right)
^{\left( N-1-n\right) }$ for $u\in\left[ 0,1\right] $. Using the
identity

\begin{equation}
\int\left( G\left( n-1,N,u\right) -G\left( n,N,u\right) \right)
du=\binom{{N}}{n}u^{n}\left( \,1-u\right) ^{\left( N-n\right) }
\label{appendixE2}
\end{equation}
we obtain

\begin{align}
D_{M}(|j,m\rangle,|j,m-1\rangle) & =2\binom{{N}}{n}\int_{0}^{\pi/2}\cos
^{2n}v\sin^{2\left(N-n\right)}vdv
\nonumber \\
&
=\binom{{N}}{n}\frac{\Gamma\left( n+1/2\right) \Gamma\left(
N-n+1/2\right)}{\Gamma\left( N\right)}
\nonumber \\
&
=\pi\binom{2\left( N-n\right) }{N-n}\binom{2n}{n}2^{-2N}\sim\frac{1}
{\sqrt{N-n}\sqrt{n}} \,
\text{,}
\label{appendixE3}
\end{align}
as desired.
	
\subsection{Derivation of formula (\ref{monostar})}

For $j\in\mathbb{N}$ we put $\ D_{j}:=D_{M}(|j,0\rangle,\rho_{\ast})$. From
Proposition 5 and formula (\ref{salv2}) we deduce that

\begin{align}
D_{j} & = \int_{0}^{\pi}\left| \int_{0}^{\vartheta}h_{|j,0\rangle
\left\langle j,0\right| }(\psi)d\psi-\int_{0}^{\vartheta}h_{\rho_{\ast}}
(\psi)d\psi\right| d\vartheta
\nonumber \\
&
=\int_{0}^{\pi}\left| \int_{0}^{\vartheta}\left( G\left( \cos^{2}\left(
\psi/2\right) \right) -1\right) \frac{1}{2}\sin\psi d\psi\right|
d\vartheta \,
\text{,}
\label{appendixE4}
\end{align}
where $G\left( u\right) :=\left( 2j+1\right) \binom{{2j}}
{j}u^{j}\left( \,1-u\right) ^{j}$ for $u\in\left[ 0,1\right] $. Applying
the substitutions $u=\cos^{2}\left( \psi/2\right) $ and $y=\vartheta/2$
and using the symmetry arguments yields

\begin{align}
\ D_{j} & = 2\int_{0}^{\pi/2}\left| \int_{\cos^{2}\left( \vartheta/2\right)
}^{1}\left( G\left( u\right) -1\right) du\right| d\vartheta
\nonumber \\
&
=\text{ }2\int_{0}^{\pi/4}\int_{\sin^{2}y}^{\cos^{2}y}G\left( u\right)
\,du\,dy-1
\nonumber \\
&
= 2\left( 2j+1\right) \binom{{2j}}{j}\int_{0}^{\pi/4}\int_{\sin^{2}y}
^{\cos^{2}y}u^{j}\left( \,1-u\right) ^{j}\,du\,dy-1 \,
\text{.}
\label{appendixE5}
\end{align}

Set $c_{j}\left( u\right) :=2\left( 2j+1\right) \binom{{2j}}{j}\int
u^{j}\left( \,1-u\right) ^{j}du$ for $u\in\left[0,1\right]$, $j\in\mathbb{N}$.
Then $D_{j}=\int_{0}^{\pi/4}\left( c_{j}\left( \cos
^{2}y\right) -c_{j}\left( \sin^{2}y\right) \right) dy-1$. Integrating by
parts we get $c_{j}\left( u\right) =2\binom{{2j}}{j}\left( 2u-1\right)
u^{j}\left( 1-u\right) ^{j}+c_{j-1}\left( u\right) $, and so
$D_{j}=\binom{{2j}}{j}2^{-2j}\frac{1}{2j+1}+D_{j-1}$. Moreover we can put $D_{0}
=0$. Thus $D_{j}=\sum_{k=1}^{j}\allowbreak\frac{1}{2k+1}2^{-2k}\binom{{2k}
}{k}=\sum_{k=1}^{j}\allowbreak\frac{1}{2k+1}\frac{\left( 2k-1\right)
!!}{\left( 2k\right) !!}$, as claimed. Applying Taylor's formula $\arcsin
x=\sum_{k=0}^{\infty}\allowbreak\frac{1}{2k+1}\frac{\left( 2k-1\right)
!!}{\left(2k\right) !!}x^{2k+1}$ we obtain $D_{j}\rightarrow\pi/2-1 \,
\left(j\rightarrow\infty\right) $.

\subsection{Derivation of formula (\ref{moncohstat})}

Let $ C(\Xi,j) = D_{M}(\rho_+,\rho_\Xi)$ for $\Xi\in\left[
0,\pi\right] $ and $j=1/2,1,...$ . It follows from the rotational invariance
of the Monge metric (Property~B) that
$C(\Xi,j) = D_{M}(\rho_1,\rho_2)$, where $\rho_{1} = \rho_{(\pi-\Xi)/2}$ and
$\rho_{2} = \rho_{(\pi+\Xi)/2}$. To apply Proposition 6 observe first that
according to formula (\ref{overlaHus}) we have

\begin{equation}
H_{\rho_{1}}(\vartheta,\varphi)=\left( {2j+1}\right) \left( \frac
{{1+\sin\vartheta\cos}\left( {\Xi/2}\right) {\cos\varphi+\cos\vartheta\sin
}\left( {\Xi/2}\right) }{2}\right) ^{2j} \text{,}
\label{E6}
\end{equation}

\begin{equation}
H_{\rho_{2}}(\vartheta,\varphi)=\left( {2j+1}\right) \left( \frac
{{1+\sin\vartheta\cos\left( {\Xi/2}\right) \cos\varphi-\cos\vartheta\sin
}\left( {\Xi/2}\right) }{2}\right) ^{2j} \text{,}
\label{E7}
\end{equation}
and so $H_{\rho_{1}}(\vartheta,\varphi)=H_{\rho_{2}}(\pi-\vartheta,\varphi)$
for $(\vartheta,\varphi)\in S^{2}$. Thus, applying the substitution
${\pi-\vartheta\rightarrow\vartheta}$, we get $F_{1}(\pi,\varphi)=\frac{1}
{2}\int_{0}^{\pi}H_{\rho_{i}}(\vartheta,\varphi)\sin\vartheta\ d\vartheta
= \frac{1}{2}\int_{0}^{\pi}H_{\rho_{2}}(\vartheta,\varphi
)\sin\vartheta\ d\vartheta=F_{2}(\pi,\varphi)$, and $F_{1}(t,\varphi
)-F_{2}(t,\varphi)=F_{1}(\pi-t,\varphi)-F_{2}(\pi-t,\varphi)$ for $t\in\left[
0,\pi\right] $ and $\varphi\in\left[ 0,2\pi\right] $, which implies the
assumption (1). Moreover,
$H_{\rho_{1}}(\vartheta,\varphi)-H_{\rho_{2}}(\vartheta,\varphi)\geq0$ for
$\vartheta\in\left[ 0,\pi/2\right]$ and $\varphi\in\left[ 0,2\pi\right]$. From
this fact and from the symmetry of the functions
$F_1(\cdot,\varphi) - F_2(\cdot,\varphi)$ ($\varphi \in \left[0,2\pi \right]$)
we deduce the assumption (2). Hence the assumptions of Proposition 6 are
fulfilled and we conclude that

\begin{equation}
C(\Xi,j)={{\frac{{2j+1}}{\pi4^{j+1}}}}\int_{0}^{2\pi}\int_{0}^{\pi}\left(
\int_{0}^{t}\left( \left( w+z\right) ^{2j}-\left( w-z\right)
^{2j}\right) {\sin\vartheta \, d\vartheta}\right) dt \, {d\varphi}
\, \text{,}
\label{E8}
\end{equation}
where $w:={1+\sin\vartheta\cos\left( {\Xi/2}\right) \cos\varphi}$ and
$z:={\cos\vartheta\sin\left( {\Xi/2}\right) }$. Applying the identity

\begin{equation}
\left( w+z\right) ^{2j}-\left( w-z\right) ^{2j}=\left\{
\begin{tabular}
[c]{ll}
$2z\sum_{k=0}^{j-1}\binom{2j}{2k+1}w^{2k+1}z^{2j-2k-2}$ & \ for $2j$ - even
\medskip \\
$2z\sum_{k=0}^{j-1/2}\binom{2j}{2k}w^{2k}z^{2j-2k-1}$ & \ for $2j$ - odd
\end{tabular}
\right. \label{E9}
\end{equation}
and performing the integration we get after tedious (but elementary) calculation
the desired result.

\bigskip


\end{document}